\documentclass[fleqn, usenatbib, useAMS]{mnras}

\usepackage{newtxtext,newtxmath}
\usepackage[T1]{fontenc}

\DeclareRobustCommand{\VAN}[3]{#2}
\let\VANthebibliography\thebibliography
\def\thebibliography{\DeclareRobustCommand{\VAN}[3]{##3}\VANthebibliography}

\usepackage{ae,aecompl}
\usepackage[utf8]{inputenc}

\usepackage{graphicx}	
\usepackage{amsmath}	
\usepackage{bm}		    
\usepackage{orcidlink}

\newcommand{\Tr}{\mathrm{Tr}}
\newcommand{\ddif}{\mathrm{d}}
\newcommand{\lya}{Ly$\alpha$\ }

\newcommand{\poned}{$P_{\mathrm{1D}}$\ }
\newcommand{\skm}{~s~km$^{-1}$\ }
\newcommand{\kms}{~km~s$^{-1}$\ }

\title[DESI Early 1D Power Spectrum]{Optimal 1D \lya Forest Power Spectrum Estimation -- III. DESI early data}

\author[N.G.~Kara{\c c}ayl{\i} et al.]{
\parbox{\textwidth}{
\Large
Naim~G{\" o}ksel~Kara{\c c}ayl{\i},$^{1,2,3}$\orcidlink{0000-0001-7336-8912}
Paul~Martini,$^{1,2}$\orcidlink{0000-0002-4279-4182}
Julien~Guy,$^{4}$
Corentin~Ravoux,$^{5,6}$\orcidlink{0000-0002-3500-6635}
Marie~Lynn~Abdul~Karim,$^{6}$
Eric~Armengaud,$^{6}$\orcidlink{0000-0001-7600-5148}
Michael~Walther,$^{7,8}$\orcidlink{0000-0002-1748-3745}
J.~Aguilar,$^{4}$
S.~Ahlen,$^{9}$
S.~Bailey,$^{4}$
J.~Bautista,$^{5}$
S.F.~Beltran,$^{10}$
D.~Brooks,$^{11}$
L.~Cabayol-Garcia,$^{12}$
S.~Chabanier,$^{4}$
E.~Chaussidon,$^{6}$
J.~Chaves-Montero,$^{12}$
K.~Dawson,$^{13}$
R.~de la Cruz,$^{10}$
A.~de la Macorra,$^{14}$
P.~Doel,$^{11}$
A.~Font-Ribera,$^{12}$
J.~E.~Forero-Romero,$^{15,16}$
S.~Gontcho A Gontcho,$^{4}$
A.X.~Gonzalez-Morales,$^{17,10}$
C.~Gordon,$^{12}$
H.K~Herrera-Alcantar,$^{10}$
K.~Honscheid,$^{1,3}$
V.~Ir\v{s}i\v{c},$^{18}$
M.~Ishak,$^{19}$
R.~Kehoe,$^{20}$
T.~Kisner,$^{4}$
A.~Kremin,$^{4}$
M.~Landriau,$^{4}$
L.~Le~Guillou,$^{21}$
M.E.~Levi,$^{4}$
Z.~Luki\'c,$^{4}$
A.~Meisner,$^{22}$
R.~Miquel,$^{23,12}$
J.~Moustakas,$^{24}$
E.~Mueller,$^{25}$
A.~Muñoz-Gutiérrez,$^{14}$
L.~Napolitano,$^{26}$
J.~Nie,$^{27}$
G.~Niz,$^{10,28}$
N.~Palanque-Delabrouille,$^{6,4}$
W.J.~Percival,$^{29,30,31}$
M.~Pieri,$^{32}$
C.~Poppett,$^{4,33}$
F.~Prada,$^{34}$
I.~P\'erez-R\`afols,$^{12}$
C.~Ram\'irez-P\'erez,$^{12}$
G.~Rossi,$^{35}$
E.~Sanchez,$^{36}$
H.~Seo,$^{37}$
F.~Sinigaglia,$^{38,39}$
T.~Tan,$^{21}$
G.~Tarl\'{e},$^{40}$
B.~Wang,$^{41}$
B.A.~Weaver,$^{22}$
C.~Yèche$^{6}$
and Z.~Zhou$^{27}$
}
\vspace{0.4cm}
\\
\parbox{\textwidth}{
$^{1}$ Center for Cosmology and AstroParticle Physics, The Ohio State University, 191 West Woodruff Avenue, Columbus, OH 43210, USA\\
$^{2}$ Department of Astronomy, The Ohio State University, 4055 McPherson Laboratory, 140 W 18th Avenue, Columbus, OH 43210, USA\\
$^{3}$ Department of Physics, The Ohio State University, 191 West Woodruff Avenue, Columbus, OH 43210, USA\\
$^{4}$ Lawrence Berkeley National Laboratory, 1 Cyclotron Road, Berkeley, CA 94720, USA\\
$^{5}$ Aix Marseille Univ, CNRS/IN2P3, CPPM, Marseille, France\\
$^{6}$ IRFU, CEA, Universit\'{e} Paris-Saclay, F-91191 Gif-sur-Yvette, France\\
$^{7}$ Excellence Cluster ORIGINS, Boltzmannstrasse 2, D-85748 Garching, Germany\\
$^{8}$ University Observatory, Faculty of Physics, Ludwig-Maximilians-Universit\"{a}t, Scheinerstr. 1, 81677 M\"{u}nchen, Germany\\
$^{9}$ Physics Dept., Boston University, 590 Commonwealth Avenue, Boston, MA 02215, USA\\
$^{10}$ Departamento de F\'{i}sica, Universidad de Guanajuato - DCI, C.P. 37150, Leon, Guanajuato, M\'{e}xico\\
$^{11}$ Department of Physics \& Astronomy, University College London, Gower Street, London, WC1E 6BT, UK\\
$^{12}$ Institut de F\'{i}sica d’Altes Energies (IFAE), The Barcelona Institute of Science and Technology, Campus UAB, 08193 Bellaterra Barcelona, Spain\\
$^{13}$ Department of Physics and Astronomy, The University of Utah, 115 South 1400 East, Salt Lake City, UT 84112, USA\\
$^{14}$ Instituto de F\'{\i}sica, Universidad Nacional Aut\'{o}noma de M\'{e}xico,  Cd. de M\'{e}xico  C.P. 04510,  M\'{e}xico\\
$^{15}$ Departamento de F\'isica, Universidad de los Andes, Cra. 1 No. 18A-10, Edificio Ip, CP 111711, Bogot\'a, Colombia\\
$^{16}$ Observatorio Astron\'omico, Universidad de los Andes, Cra. 1 No. 18A-10, Edificio H, CP 111711 Bogot\'a, Colombia\\
$^{17}$ Consejo Nacional de Ciencia y Tecnolog\'{\i}a, Av. Insurgentes Sur 1582. Colonia Cr\'{e}dito Constructor, Del. Benito Ju\'{a}rez C.P. 03940, M\'{e}xico D.F. M\'{e}xico\\
$^{18}$ Kavli Institute for Cosmology, University of Cambridge, Madingley Road, Cambridge CB3 0HA, UK\\
$^{19}$ Department of Physics, The University of Texas at Dallas, Richardson, TX 75080, USA\\
$^{20}$ Department of Physics, Southern Methodist University, 3215 Daniel Avenue, Dallas, TX 75275, USA\\
$^{21}$ Sorbonne Universit\'{e}, CNRS/IN2P3, Laboratoire de Physique Nucl\'{e}aire et de Hautes Energies (LPNHE), FR-75005 Paris, France\\
$^{22}$ NSF's NOIRLab, 950 N. Cherry Ave., Tucson, AZ 85719, USA\\
$^{23}$ Instituci\'{o} Catalana de Recerca i Estudis Avan\c{c}ats, Passeig de Llu\'{\i}s Companys, 23, 08010 Barcelona, Spain\\
$^{24}$ Department of Physics and Astronomy, Siena College, 515 Loudon Road, Loudonville, NY 12211, USA\\
$^{25}$ Institute of Cosmology \& Gravitation, University of Portsmouth, Dennis Sciama Building, Portsmouth, PO1 3FX, UK\\
$^{26}$ Department of Physics \& Astronomy, University  of Wyoming, 1000 E. University, Dept.~3905, Laramie, WY 82071, USA\\
$^{27}$ National Astronomical Observatories, Chinese Academy of Sciences, A20 Datun Rd., Chaoyang District, Beijing, 100012, P.R. China\\
$^{28}$ Instituto Avanzado de Cosmolog\'{\i}a A.~C. San Marcos 11 - Atenas 202. Magdalena Contreras, 10720. Ciudad de M\'{e}xico, M\'{e}xico\\
$^{29}$ Department of Physics and Astronomy, University of Waterloo, 200 University Ave W, Waterloo, ON N2L 3G1, Canada\\
$^{30}$ Perimeter Institute for Theoretical Physics, 31 Caroline St. North, Waterloo, ON N2L 2Y5, Canada\\
$^{31}$ Waterloo Centre for Astrophysics, University of Waterloo, 200 University Ave W, Waterloo, ON N2L 3G1, Canada\\
$^{32}$ Aix Marseille Univ, CNRS, CNES, LAM, Marseille, France\\
$^{33}$ Space Sciences Laboratory, University of California, Berkeley, 7 Gauss Way, Berkeley, CA  94720, USA\\
$^{34}$ Instituto de Astrof\'{i}sica de Andaluc\'{i}a (CSIC), Glorieta de la Astronom\'{i}a, s/n, E-18008 Granada, Spain\\
$^{35}$ Department of Physics and Astronomy, Sejong University, Seoul, 143-747, Korea\\
$^{36}$ CIEMAT, Avenida Complutense 40, E-28040 Madrid, Spain\\
$^{37}$ Department of Physics \& Astronomy, Ohio University, Athens, OH 45701, USA\\
$^{38}$ Instituto de Astrof\'{i}sica de Canarias, C/ Vía L\'{a}ctea, s/n, E-38205 La Laguna, Tenerife, Spain\\
$^{39}$ Universidad de La Laguna, Dept. de Astrof\'{\i}sica, E-38206 La Laguna, Tenerife, Spain\\
$^{40}$ Department of Physics, University of Michigan, Ann Arbor, MI 48109, USA\\
$^{41}$ Department of Astronomy, Tsinghua University, 30 Shuangqing Road, Haidian District, Beijing, China, 100190
}
}

\date{Accepted XXX. Received YYY; in original form ZZZ}

\pubyear{2022}

\begin{document}
\label{firstpage}
\pagerange{\pageref{firstpage}--\pageref{lastpage}}
\maketitle

\begin{abstract}
The one-dimensional power spectrum \poned of the \lya forest provides important information about cosmological and astrophysical parameters, including constraints on warm dark matter models, the sum of the masses of the three neutrino species, and the thermal state of the intergalactic medium. We present the first measurement of \poned with the quadratic maximum likelihood estimator (QMLE) from the Dark Energy Spectroscopic Instrument (DESI) survey early data sample. This early sample of $54~600$ quasars is already comparable in size to the largest previous studies, and we conduct a thorough investigation of numerous instrumental and analysis systematic errors to evaluate their impact on DESI data with QMLE. We demonstrate the excellent performance of the spectroscopic pipeline noise estimation and the impressive accuracy of the spectrograph resolution matrix with two-dimensional image simulations of raw DESI images that we processed with the DESI spectroscopic pipeline. We also study metal line contamination and noise calibration systematics with quasar spectra on the red side of the \lya emission line. In a companion paper, we present a similar analysis based on the Fast Fourier Transform estimate of the power spectrum. We conclude with a comparison of these two approaches and discuss the key sources of systematic error that we need to address with the upcoming DESI Year 1 analysis. 
\end{abstract}
\begin{keywords}
methods: data analysis -- intergalactic medium -- quasars: absorption lines
\end{keywords}



\section{Introduction}
\begin{figure*}
    \centering
    \includegraphics[width=\linewidth]{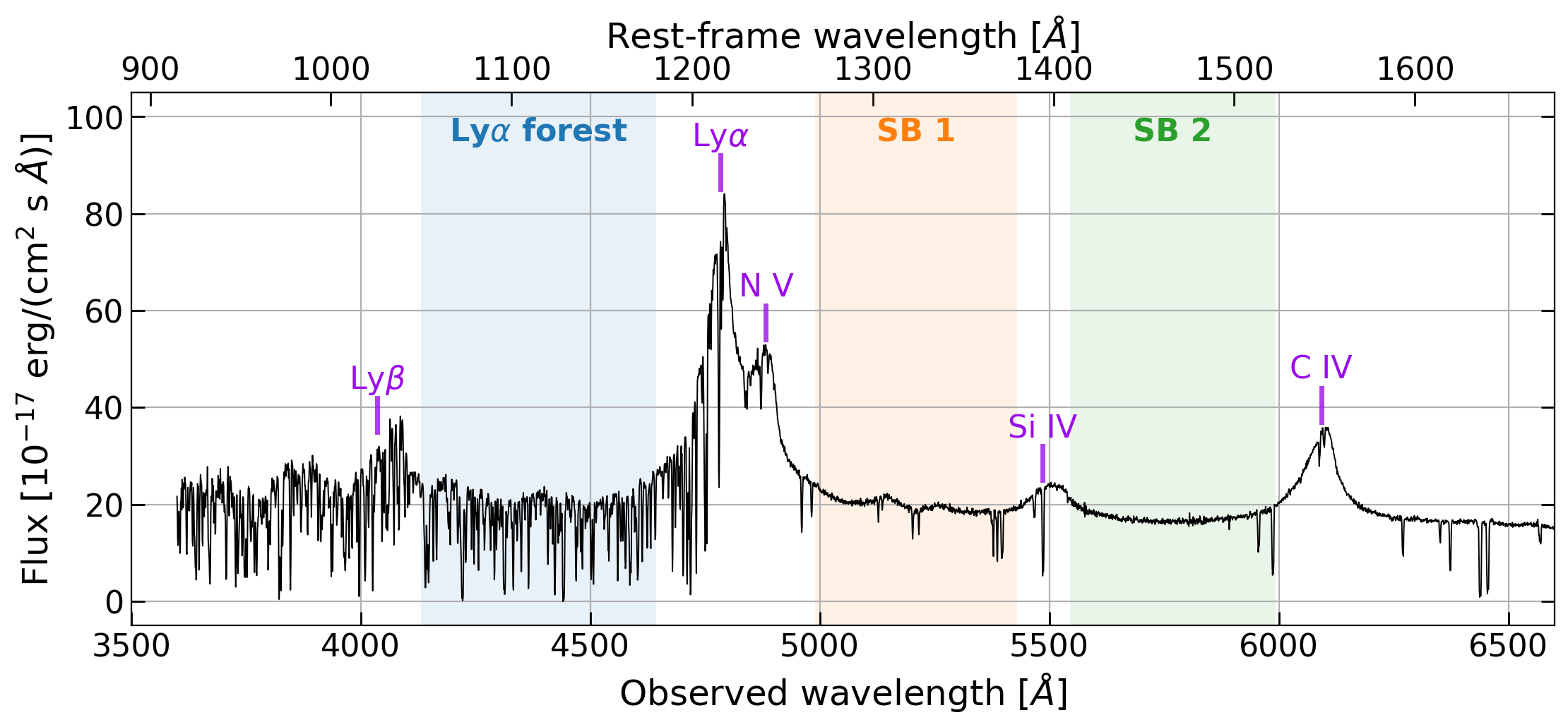}
    \caption{Quasar at $z=2.94$ observed during DESI survey validation (TargetID 39627871806818826).
    The \lya forest is defined to be the spectral region between a quasar's \lya and \ion{Ly}{$\beta$} emission lines.
    Absorption features redward of the quasar's \lya emission line may be due to metal systems.
    The regions from \lya to \ion{Si}{IV} and from \ion{Si}{IV} to \ion{C}{IV} are called the "side bands" (SB).
    We call the \lya--\ion{Si}{IV} region SB~1 and the \ion{Si}{IV}--\ion{C}{IV} region SB~2, and use these regions to quantify metal contamination, noise and flux calibrations.
    }
    \label{fig:fuji_example_spectrum}
\end{figure*}
Neutral hydrogen gas between us and distant quasars forms absorption lines at wavelengths shorter than the \lya emission line in the quasar spectrum through absorption and scattering.
These absorption lines are collectively called the \lya forest; and they trace the underlying matter distribution in the intergalactic medium (IGM) and the circumgalactic medium (CGM).
The \lya forest is consequently a powerful tool to map vast volumes at redshifts $2\lesssim z \lesssim 5$ and probing scales from hundreds of Mpc to below 1~Mpc.

\citet{gunnDensityNeutralHydrogen1965} first estimated the density of neutral hydrogen in the IGM. They realised that the measurement of some continuum flux of 3C~9 below the \lya emission line by \citet{schmidtQuasarRedshifts1965} implied the IGM was mostly ionized.
Later work by \citet{lyndsAbsorptionLineSpectrum4c1971} showed that the IGM absorption was in the form of discrete features. In the 1990s, work by many investigators \citep{biAlternativeModelLyalpha1992, cenGravitationalCollapseSmallscale1994, zhangMultispeciesModelHydrogen1995, biEvolutionStructureIntergalactic1997} clearly established that this \lya forest originates from smooth IGM fluctuations.
Based on this smooth density fluctuations picture, the 1D power spectrum (\poned) has emerged as an important quantity to measure in high-resolution, high-signal-to-noise (SNR) spectra \citep{croftRecoveryPowerSpectrum1998, irsicLymanEnsuremathAlpha2017, waltherNewPrecisionMeasurement2017, karacayliOptimal1DLy2022}, as well as medium-resolution, medium-SNR spectra \citep{mcdonaldLyUpalphaForest2006, palanque-delabrouilleOnedimensionalLyalphaForest2013, chabanierOnedimensionalPowerSpectrum2019}. 
\poned is valuable because it is sensitive to smaller scales than are accessible in high-redshift galaxy surveys, and consequently to particular physical quantities. Applications of the \lya \poned include  
investigations of the thermal state of the IGM \citep{boeraRevealingThermal2019, waltherNewConstraintsIGM2019, villasenorThermalHistory2022},
inference of the primordial power spectrum \citep{vielPrimordialPowerSpectrumLya2004},
constraints on the sum of neutrino masses \citep{croftNeutrinoMassLyaForest1999, palanqueDelabrouilleNeutrinoMass2015, yecheNeutrinoMassesXQ2017},
and explorations of the nature of dark matter \citep{narayananWDMLyaForest2000, seljakSterileNeutrinosDM2006, wangLyaDecayingDM2013, irsicFuzzyDMfromLya2017}, with warm dark matter receiving particular attention \citep{boyarskyLyaWDM2009, vielWarmDarkMatter2013, baurLyaCoolWDM2016, irsicConstraintsWDM2017, villasenorWDM2022}. 

Even though \poned is a summary statistic for cosmological analysis, it is very sensitive to several sources of systematic errors. 
The five-year data from the Dark Energy Spectroscopic Instrument \citep[DESI,][]{leviDESIExperimentWhitepaper2013} will provide approximately 700\,000 \lya quasar spectra with medium resolution ($R\approx 3~000$), medium signal-to-noise (SNR) ($\approx 2$ per \AA) \citep{desicollaborationDESIExperimentPart2016, desicollaborationDESIExperimentPart2016b}, which will constitute a data set that is four times larger than the Extended Baryon Oscillation Spectroscopic Survey \citep[eBOSS,][]{dawsonSDSSeBOSSEarlyData2016}.
DESI will consequently substantially expand the statistical power of \lya forest measurements relative to previous work. 
To fully exploit this great increase in statistical power requires comprehensive studies of \poned systematics.  These include systematics related to the theoretical interpretation \citep[e.g.,][]{lukicLyaOpticallyThinSimulations2015, waltherSimulatingIGDESI2020, chabanierModellingLya2023}, instrumental effects, and other spectroscopic extraction details. We address the latter two topics in this paper with early DESI observations.  

We analyze two distinct datasets in this paper. The first set of spectra was collected between December 2020 and May 2021 during the DESI Survey Validation \citep[SV,][]{surveyValidation2023} phase.
The purpose of this phase was to perform various tests to verify the pipeline for target selection, spectral extraction, classifier performance and clustering analysis. The spectra collected during this period will be publicly available as early data release \citep{earlyDataRelease2023}.
The second set was obtained during the first two months of the DESI main survey, which began in May 2021. 
Together, these data span a wide range of signal-to-noise ratios (SNR). We use them to measure \poned and characterize the noise, flux calibration, and spectrograph resolution calculated by the DESI spectroscopic pipeline.

The two main methods to estimate \poned are the maximum likelihood estimator and the Fast Fourier transform (FFT). 
The maximum likelihood estimator is typically considered to be statistically optimal, although it is slower than FFT-based algorithms.
The maximum likelihood estimator can be implemented in two different ways.
A direct implementation finds the maximum likelihood solution by sampling the likelihood with respect to \poned \citep{palanque-delabrouilleOnedimensionalLyalphaForest2013}.
This implementation has slower convergence properties and is more prone to numerical instabilities.
The second implementation takes advantage of the Newton-Raphson method and achieves a faster and more stable performance.
We call this estimator the quadratic maximum likelihood estimator \citep[QMLE;][]{mcdonaldLyUpalphaForest2006, fontriberaEstimate3DPowerLya2018, karacayliOptimal1DLy2020} and the application of QMLE to DESI data is the main focus of this paper. In a companion paper by \citet{ravouxFFTP1dEDR2023} we present the application of the FFT-based estimator to early DESI data. That paper adapts the FFT approach previously used for eBOSS \citep{chabanierOnedimensionalPowerSpectrum2019}.

A major virtue of QMLE is that it is robust against challenges such as strong sky emission lines, high-column density (HCD) systems, and bad CCD pixels. Pixels affected by these features must be masked to avoid contamination from unrelated physical effects and imperfections in instrumentation.
This masking introduces a bias that must be corrected in FFT estimates; and these corrections in turn introduce uncertainties to the measurement \citep{chabanierOnedimensionalPowerSpectrum2019}.
A major advantage of QMLE is that it can handle masked, uneven spectra without further corrections by construction.
Relatedly, QMLE is capable of weighting individual pixels by the inverse pipeline noise, and hence diminishes the impact of variations in instrument noise and other noisy spectral regions such as certain sky lines.
In addition, the QMLE implementation of \citet{karacayliOptimal1DLy2020} interpolates pixel pairs into two redshift bins to account for the redshift evolution within the \lya forest.
These properties, among others discussed later in the text, make the QMLE an excellent tool for DESI \lya \poned estimation.

For a medium-resolution, medium-SNR survey such as DESI, the potential systematics due to the pipeline noise estimation and the spectrograph resolution require the most attention.
Previous experiments suffered from spectroscopic pipeline noise miscalibration levels of $15\%$, which necessitated separate calculations and recalibrations of the pipeline noise \citep{mcdonaldLyUpalphaForest2006, palanque-delabrouilleOnedimensionalLyalphaForest2013}.
DESI was meticulously designed to abate such miscalibrations \citep{abareshiOverviewInstrumentationDark2022, guySpectroscopicDataProcessingPipeline2022}.
Yet even though the pipeline is significantly improved, the statistical power of even the early data demands ever-stringent precision.
Another consideration is that spectral extraction for DESI is based on the spectro-perfectionism algorithm, which can handle arbitrarily complicated (i.e. not solely separable) 2D point-spread functions (PSF) \citep{boltonSpectroPerfectionismAlgorithmicFramework2010}.
This extraction preserves the full native resolution of the 2D spectrograph without degradation in the 1D spectrum and yields an independent resolution matrix for each 1D spectrum that is based on the spectrograph resolution and the noise in each spectrum \citep{guySpectroscopicDataProcessingPipeline2022}.
QMLE can naturally incorporate this novel resolution matrix, and in this paper, we validate the spectro-perfectionism and its synergy with the QMLE by simulating CCD images and extracting spectra with the DESI spectroscopic pipeline.

The outline of this paper is as follows. First, we describe the DESI survey, target selection, the creation of quasar catalogues, the identification of damped \lya (DLA) systems and broad absorption lines (BAL), and the properties of the early spectra in Section~\ref{sec:data}.
We outline the continuum fitting algorithm and detail the QMLE and various updates in  Section~\ref{sec:method}.
Synthetic spectra are central in our validation to make robust statistical claims.
In Section~\ref{sec:validation}, we validate the continuum fitting algorithm, DLA masking and damping wing corrections with extensive sets of 1D mock spectra, and validate the resolution matrix derived by the pipeline CCD image simulations that we analyze with the same spectroscopic pipeline that we use with real DESI observations.
We perform various tests for systematics and present our \poned measurement from data in Section~\ref{sec:results}. Finally, we compare DESI \poned measurements from the QMLE and FFT estimators to each other and to eBOSS in Section~\ref{sec:discussion}.
We summarize our results in Section~\ref{sec:summary}. As noted before, a companion paper by \citet{ravouxFFTP1dEDR2023} presents the FFT-based results.

\section{Data}
\label{sec:data}
The DESI collaboration began a five-year survey of 40 million galaxies and quasars in May 2021. The main goal of this survey is to measure distances with the Baryon Acoustic Oscillation (BAO) method from the local universe to beyond $z > 3.5$ and use these data to explore the nature of dark energy. DESI will also employ Redshift Space Distortions to measure the growth of cosmic structures and test potential modifications to general relativity, measure the sum of neutrino masses, and investigate primordial density fluctuations from the inflationary epoch. The collaboration is conducting this survey with a new, high-throughput, fiber-fed spectrograph on the 4\,m Mayall telescope that can obtain 5000 spectra in each observation \citep{desicollaborationDESIExperimentPart2016b, silberRoboticMultiobjectFocal2023}. The light from each fiber is directed into one of ten, identical, bench-mounted spectrographs that record the light from $360-980$\,nm in three wavelength channels. The blue channel is optimized for \lya forest studies and extends from $360-593$\,nm with a resolution that ranges from $2000-3500$. These spectrographs are in a climate-controlled enclosure that provides very stable calibrations and minimizes systematic errors due to instrumental effects. The instrumentation is described in detail in \citet{abareshiOverviewInstrumentationDark2022} and the spectroscopic pipeline in \citet{guySpectroscopicDataProcessingPipeline2022}.

DESI targets were selected with $g, r, z$ photometry from the Legacy Imaging Surveys \citep{deyOverviewDESILegacy2019} and $W1, W2$ photometry from the Wide-field Infrared Explorer \citep[WISE,][]{wrightWideFieldWISE2010}. The target selection process is described in detail in \citet{myersTargetSelectionPipelineDESI2022}. The targets include quasars at $0.9 < z < 2.1$ that are used to trace large-scale structure and at $z > 2.1$ that are used to trace the matter distribution with the \lya forest \citep{yechePreliminaryTargetQSODESI2020}. The collaboration refined the target selection algorithms during the Survey Validation \citep[SV,][]{surveyValidation2023} period in early 2022 with a significant visual inspection effort \citep{alexanderDESISVVIQSO2022}. The final quasar target selection is based on a random forest algorithm and selects quasars in the magnitude range $16.5<r<23$ \citep{chaussidonTargetSelectionDESIQSO2022}.
We use the One-Percent Survey (SV3) spectra that are part of early data release \citep[EDR,][]{earlyDataRelease2023}, and further include two months of main survey (DESI-M2) to increase the statistical precision in our analysis. We call this combined data set EDR+. The Target Selection Validation (SV1) spectra are the deepest observations in EDR, but their pipeline noise estimates differ from the other two data sets \citep{ravouxFFTP1dEDR2023}. Therefore, we rely on these spectra only for the DLA identification and not for \poned estimation since the pipeline noise estimates do not affect DLA identification as they affect \poned.
Figure~\ref{fig:fuji_example_spectrum} shows a quasar at $z=2.94$ from this DESI early data.

DESI employs three classification algorithms to identify quasars. Most targets are correctly classified with Redrock\footnote{\url{https://github.com/desihub/redrock}} \citep[in preparation]{baileyRedrock2023}. This algorithm performs a $\chi^2$ analysis for a range of spectral templates as a function of redshift and identifies the best redshift and spectral template for each target. Our visual inspection process demonstrated that Redrock misses some quasars, so we employ QuasarNET \citep{buscaQuasarNET2018, farrQuasarNetDESI2020} and an \ion{Mg}{II} afterburner \citep{napolitanoMgIIAfterburner2023} to help identify additional quasars. QuasarNET is a machine learning algorithm that uses convolutional neural networks for classification and the \ion{Mg}{II} afterburner searches for broad \ion{Mg}{II} emission at the Redrock redshift in the spectral of quasar targets classified as galaxies. \citet{chaussidonTargetSelectionDESIQSO2022} describe this process in more detail. 
We limit ourselves to objects that are targeted as quasars in the afterburner catalogue.

\subsection{Quasars with broad absorption lines}
Broad absorption line (BAL) features are present in about 15\% of all quasar spectra and can contaminate the \lya forest as well as impact quasar redshift errors and classifications. The vast majority of BAL quasars exhibit blueshifted absorption associated with the \ion{C}{IV} emission feature and the BAL identification algorithm searches this region in every quasar spectrum where this spectral region is visible ($1.57 < z < 5$). This algorithm is similar to the one presented by \citet{guoClassificationBALs2019}, except that it does not use the Convolutional Neural Network (CNN) classifier. \citet{filbertBALEDRcatalog2023} describe the BAL identification and characterization for the early DESI quasar catalogues in detail, including the catalogue completeness and purity, and the impact of BAL features on redshift errors \citep{garciaImpactBALRedshiftErrors2023}. We use the measured velocity range of the BAL features associated with \ion{C}{IV} to mask this ion and also mask the wavelengths that correspond to the same velocities associated with the \ion{S}{IV}, \ion{P}{V}, \ion{C}{III}, \lya\!, \ion{N}{V}, and \ion{Si}{IV}. All of these features may be present in BALs \citep{masribasBALs2019}, and all but \ion{S}{IV} can contaminate the \lya forest \citep[e.g.,][]{ennesserMitigationBALquasars2022}.

\subsection{Damped \lya systems}
DLAs are identified using both Convolutional Neural Network (CNN) and Gaussian process (GP) finders, then their results are combined into a concordance catalogue while adopting GP results over CNN if both detect the same DLA \citep{mingfengDLAGP2021, wangDeepLearningDESIDLA2022}.
We pick unique DLA identifications while combining three separate DLA catalogues for SV1, SV3 and DESI-M2 since the same quasars and DLAs can be present in different catalogues.
If two DLAs are within a threshold redshift separation $\Delta z_t$ that corresponds to a DLA's observed redshift size, we pick the highest confidence identification, where 
\begin{equation}
    \Delta z_t = (1+z_\mathrm{DLA}) \frac{7.3 \text{\,\AA}}{\lambda_{\mathrm{Ly\alpha}}} 10^{(N_\mathrm{HI}-20)/2}.
\end{equation}
We select systems based on average signal-to-noise ratio $\overline{\mathrm{SNR}}$ between 1420--1480~\AA\, in quasar's rest frame.
For DLAs that are identified by CNN, we keep them in the catalogue if the host quasar spectrum has $\overline{\mathrm{SNR}}>3$, but remove them from the catalogue if the confidence level is less than 0.3 in quasars with $\overline{\mathrm{SNR}}<3$.
We keep all systems GP identifies.
There are 41\,946 DLAs with $N_\mathrm{HI}>20.3$ in the combined catalogue.
Sub-DLA detections contain many false positives, so we do not mask them. 
Our selection criteria and duplicate removal reduce this number to 30\,131.
Introducing a minor confidence threshold of 0.2 for high $\overline{\mathrm{SNR}}$ and 0.9 for GP systems removes 567 DLAs.
We believe masking possible DLAs in this small sample is more valuable than missing them.
We note that not all DLA sightlines end up in our final sample since some host quasars are left out due to quality cuts.

\subsection{Redshift distribution}
\begin{figure}
    \centering
    \includegraphics[width=\columnwidth]{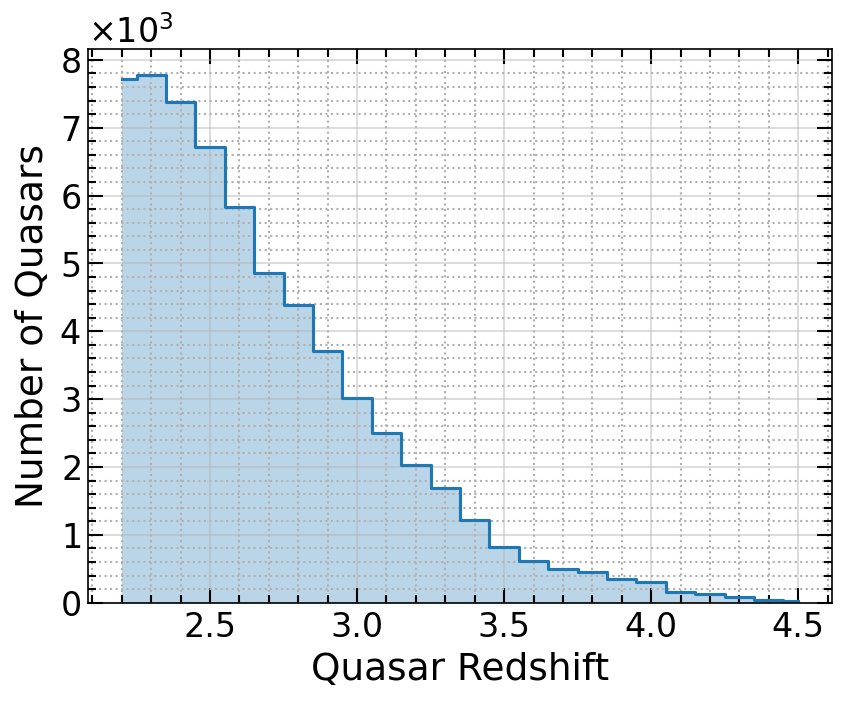}\\
    \includegraphics[width=\columnwidth]{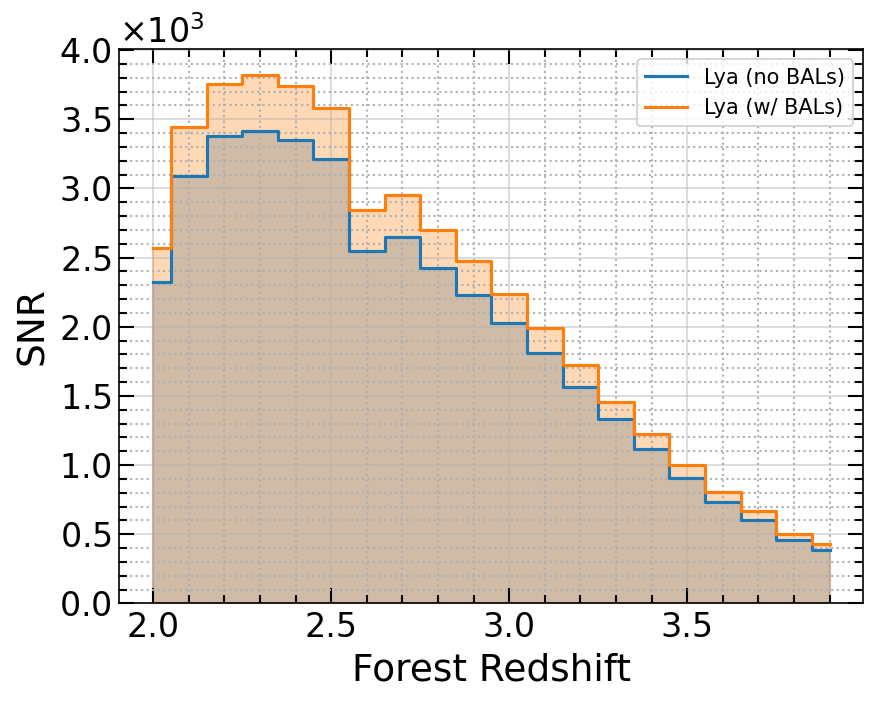}
    \caption{({\it Top}) Quasar redshift distribution of our sample.
    ({\it Bottom}) Weighted mean SNR distribution in the forest as a function of redshift, where we define SNR to be $1/\sigma(z)$ and $\sigma(z)$ is the propagated error on the weighted mean of pixel size of $\Delta z=0.1$.
    Having BAL quasars ({\it orange}) improves SNR, but it also comes with possible biases in \poned (see Section~\ref{sec:data}).
    }
    \label{fig:fujilupe-snr-nzqso}
\end{figure}

Figure~\ref{fig:fujilupe-snr-nzqso} shows the quasar redshift distribution of our sample on the top panel.
The quasar distribution $n_\mathrm{qso}(z)$ drops off rapidly at higher redshift, as expected from the selection function.
There are 67\,241 quasars in our final sample.
On the bottom panel, we show the SNR distribution in the forest as a function of redshift with bin size $\Delta z = 0.1$.
We define SNR based on the propagated error $\sigma(z)$, where SNR=$1/\sigma(z)$ and the propagated error $\sigma(z)$ on the weighted mean as follows:
\begin{equation}
    \sigma(z) = \left.\sqrt{\sum_i w_i^2 \sigma^2_{\mathrm{pipe},i}} \right/ \sum_i w_i,
\end{equation}
where $w_i^{-1} = \sigma^2_{\mathrm{LSS}, i} + \sigma^2_{\mathrm{pipe}, i}$ and summation is done over all pixels that fall into the redshift bin.
This quantity is equivalent to pixel SNR values after coadding all quasar spectra in the forest region into coarse $\Delta z = 0.1$ pixels.
Large-scale structure variance $\sigma^2_\mathrm{LSS}$ is calculated during the continuum fitting process as described in Section~\ref{sec:method}.
Even though we keep BAL quasars while continuum fitting, we remove them from final \poned estimates.
Removing these BAL quasars leaves us with 54\,600 spectra and reduces our SNR as shown in blue in Figure~\ref{fig:fujilupe-snr-nzqso}.

We explain the details regarding DLA and BAL masking in Section~\ref{sec:results}.

\section{Method}
\label{sec:method}
\subsection{Continuum fitting}
The continuum fitting algorithm we use was developed over the last few years and has been applied to both 3D analyses \citep{bautistajuliane.MeasurementBaryonAcoustic2017, bourbouxExtendedBaryonOscillation2019, bourbouxCompletedSDSSIVExtended2020} and \poned measurements \citep{chabanierOnedimensionalPowerSpectrum2019}.
This algorithm is part of the software Package for Igm Cosmological-Correlations Analyses (\textsc{picca}) and is publicly available\footnote{\url{https://github.com/igmhub/picca}}.
We summarize the algorithm below.

One important aspect of the algorithm is that the definition of the quasar continuum absorbs the mean transmission $\overline{F}(z)$ of the IGM. Specifically, we model every quasar "continuum" $\overline{F} C_q(\lambda_\mathrm{RF})$ by a global mean continuum $\overline{C}(\lambda_\mathrm{RF})$ and two quasar "diversity" parameters, amplitude $a_q$ and slope $b_q$:
\begin{align}
    \overline{F}C_q(\lambda_\mathrm{RF}) &= \overline{C}(\lambda_\mathrm{RF}) \left( a_q + b_q \Lambda \right) \\
    \Lambda &= \frac{\log\lambda_\mathrm{RF} - \log\lambda_\mathrm{RF}^{(1)}}{\log\lambda_\mathrm{RF}^{(2)} - \log\lambda_\mathrm{RF}^{(1)}}
\end{align}
where $\lambda_\mathrm{RF}$ is the wavelength in quasar's rest frame and $\lambda_\mathrm{RF}^{(1, 2)}$ are the minimum and maximum wavelengths considered for calculation.
We assume that the global mean continuum $\overline{C}(\lambda_\mathrm{RF})$ does not depend on redshift, and therefore our model only adjusts $\overline{F}(z)$, as well as solves for the $a_q$ and $b_q$ parameters for each quasar.
In other words, the amplitude and slope parameters do not only fit for intrinsic quasar diversity such as brightness, but also for the IGM mean transmission.
Given these definitions, transmitted flux fluctuations are given by
\begin{equation}
    \delta_q(\lambda) = \frac{f_q(\lambda)}{\overline{F}C_q(\lambda)} - 1,
\end{equation}
where $\lambda=(1+z_q)\lambda_\mathrm{RF}$\ is the observed wavelength and $f_q(\lambda)$ is the observed flux.
The effect of spectrograph resolution has been ignored for simplicity as noted in \citet{slosarMeasurementBaryonAcoustic2013}, since the affected scales are small for 3D analysis.
The features in the continuum are also wider than the spectrograph resolution,
so this assumption should also hold for \poned.

Our continuum fitting procedure calculates $a_q$ and $b_q$ for each quasar, and three global functions: the mean quasar continuum $\overline{C}(\lambda_\mathrm{RF})$, the large-scale \lya fluctuations $\sigma^2_\mathrm{LSS}(\lambda)$, and the pipeline noise correction term $\eta(\lambda)$. We do not assume a functional form for these three functions; instead, we construct linear interpolations based on binned estimates. Specifically, $\overline{C}(\lambda_\mathrm{RF})$ is calculated between rest-frame wavelengths $\lambda_\mathrm{RF}^{(1)}$ and $\lambda_\mathrm{RF}^{(2)}$ in bins of size $\Delta\lambda_\mathrm{RF}$. The other two parameters $\eta(\lambda)$ and $\sigma^2_\mathrm{LSS}(\lambda)$ are calculated in the observed frame in $N_\mathrm{obs}$ bins linearly spaced between $\lambda^{(1)}$ and $\lambda^{(2)}$. These binning parameters are tuned for each analysis depending on the available statistics. 
Before we start our fitting process, we co-add the three spectrograph arms using the pipeline inverse variance as weights. Our fitting procedure is iterative. Each iteration $i$ is as follows:
\begin{itemize}
    \item Fit each spectrum for $a_q$ and $b_q$ while keeping other parameters fixed.
    \item Calculate $\overline{C}_{i+1}(\lambda_\mathrm{RF})$.
    \item Fit for variance parameters $\eta$ and $\sigma^2_\mathrm{LSS}$ (defined below) for each bin.
\end{itemize}

For each quasar, we find the $a_q$ and $b_q$ values that minimize the following cost function while keeping \textit{all} other parameters fixed:
\begin{equation}
    \chi^2 = \sum_j \frac{\left[f_j - ( a_q + b_q \Lambda_j) \overline{C}\left(\frac{\lambda_j}{1 + z_q}\right) \right]^2}{\sigma_{q, j}^2} + \sum_j \ln \sigma_{q, j}^2,
\end{equation}
where the summation $j$ is over all pixels in the forest region and $\lambda_j$ is the observed wavelength.
The major complication comes from $\sigma_{q, j}^2$, which must take into account the intrinsic large-scale \lya fluctuations $\sigma^2_\mathrm{LSS}$:
\begin{equation}
    \sigma_{q, j}^2 = \eta(\lambda_j) \sigma^2_\mathrm{pipe, j} + \sigma^2_\mathrm{LSS}(\lambda_j) ( a_q + b_q \Lambda_j)^2 \overline{C}^2\left(\frac{\lambda_j}{1 + z_q}\right) \label{eq:sigma2_cfit}. 
\end{equation}
After every quasar is fit, we stack all continua in the rest frame and update the global mean continuum  $\overline{C}$. As described above, parameters $\eta$ and $\sigma^2_\mathrm{LSS}$ are calculated at discrete wavelength bins. For each bin, we rebin the $\delta$ values with respect to the pipeline noise estimates $\sigma_\mathrm{pipe}$ and calculate the scatter in these $\sigma_\mathrm{pipe}$ bins to measure the $\sigma_q^2 - \sigma_\mathrm{pipe}^2$ relation from the data.
Lastly, we fit equation~\ref{eq:sigma2_cfit} to this relation to find $\eta$ and $\sigma^2_\mathrm{LSS}$ values for every wavelength bin.

\subsection{Quadratic estimator}
We measure \poned using the quadratic maximum likelihood estimator (QMLE), which was extensively studied in the 90s in the context of cosmic microwave background radiation, galaxy surveys and weak lensing \citep{hamiltonOptimalMeasurementPower1997, tegmarkKarhunenLoeveEigenvalueProblems1997, tegmarkMeasuringGalaxyPower1998, seljakWeakLensingReconstruction1998}, and later also applied to the \lya forest \citep{mcdonaldLyUpalphaForest2006, karacayliOptimal1DLy2020, karacayliOptimal1DLy2022}.
The QMLE works in real space (instead of Fourier space) to estimate the power spectrum, and therefore allows weighting by the pipeline noise, accounts for intrinsic \lya large-scale structure correlations, and most importantly is not biased by gaps in the spectra.
We refer the reader to \citet{karacayliOptimal1DLy2020} and \citet{karacayliOptimal1DLy2022} for our development process and application to high-resolution spectra.
In this section, we provide a short summary of QMLE and then describe the important steps for the resolution matrix and shifting Nyquist frequency implementations. Details regarding the continuum error marginalization are in Appendix~\ref{app:continuum_marg} and signal-noise coupling correction is in Appendix~\ref{app:signal_noise_coupling}.

One motivation for the development of QMLE is that the power spectrum is typically estimated on discrete wavenumbers $k$ as band powers, since it cannot be estimated continuously on $k$, and this discretization inevitably averages the underlying power over these bands.
Our QMLE implementation alleviates this effect by estimating deviations from a fiducial power spectrum such that $P(k, z) = P_{\mathrm{fid}}(k, z) + \sum_{m,n} w_{(mn)}(k, z) \theta_{(mn)}$, where we adopt top-hat $k$ bands with $k_n$ as bin edges and linear interpolation for $z$ bins with $z_m$ as bin centres.
This fiducial power spectrum further improves the weighting by including large-scale \lya correlations,
does not have to exactly match the true power spectrum, and can be approximated in an unbiased way if no safe guess is available \citep[as shown by][]{karacayliOptimal1DLy2020}.
We use the following fitting function:
\begin{equation}
    \label{eq:pd13_fitting_fn}\frac{kP_{\mathrm{fid}}(k, z)}{\pi} = A \frac{(k/k_0)^{3 +n + \alpha\ln k/k_0}}{1+(k/k_1)^2} \left(\frac{1+z}{1+z_{0}}\right)^{B + \beta\ln k/k_0},
\end{equation}
where $k_{0} = 0.009$\skm and $z_{0}=3.0$, and
stress that this is sufficient for a baseline estimate, which in turn can be used to weight pixels, but does not capture all of the scientific information in \poned.

Given a collection of pixels representing normalized flux fluctuations $\bm{\delta}_F$, the quadratic estimator is formulated as follows:
\begin{equation}
    \label{eq:theta_it_est}\hat \theta^{(X+1)}_{\alpha} = \sum_{\alpha'} \frac{1}{2} F^{-1}_{\alpha\alpha'}(d_{\alpha'} - b_{\alpha'} - t_{\alpha'}),
\end{equation}
where $X$ is the iteration number and 
\begin{align}
    \label{eq:data_dn} d_{\alpha} &= \bm{\delta}_F^\mathrm{T} \mathbf{C}^{-1}\mathbf{Q}_{{\alpha}} \mathbf{C}^{-1} \bm{\delta}_F, \\
    \label{eq:noise_bn}b_{\alpha} &= \Tr(\mathbf{C}^{-1}\mathbf{Q}_{\alpha} \mathbf{C}^{-1}\mathbf{N}), \\
    \label{eq:signalfid_tn}t_{\alpha} &= \Tr(\mathbf{C}^{-1}\mathbf{Q}_{\alpha} \mathbf{C}^{-1}\mathbf{S}_{\mathrm{fid}}).
\end{align}
The covariance matrix $\mathbf C \equiv \langle\bm{\delta}_F\bm{\delta}_F^T\rangle$ is the sum of signal and noise, $\mathbf C = \mathbf{S}_{\mathrm{fid}} + \sum_{\alpha} \mathbf{Q}_{\alpha} \theta_{\alpha}+\mathbf{N}$, $\mathbf{Q}_{\alpha} = \partial \mathbf{C} / \partial \theta_{\alpha}$ and the estimate of the Fisher matrix is
\begin{equation}
    \label{eq:fisher_matrix}F_{\alpha\alpha'} = \frac{1}{2} \Tr(\mathbf{C}^{-1}\mathbf{Q}_{\alpha} \mathbf{C}^{-1} \mathbf{Q}_{\alpha'} ).
\end{equation}
The covariance matrices on the right hand side of equation~(\ref{eq:theta_it_est}) are computed using parameters from the previous iteration $\theta_{\alpha}^{(X)}$. Assuming different quasar spectra are uncorrelated, the Fisher matrix $F_{\alpha\alpha'}$ and the expression in parentheses in equation~(\ref{eq:theta_it_est}) can be computed for each quasar, then accumulated, i.e. $\mathbf{F}=\sum_q\mathbf{F}_{q}$ etc. 

We convert wavelength to velocity using logarithmic spacing, following the convention in cosmology:
\begin{align}
    v_i &= c \ln (\lambda_i/\lambda_{\mathrm{Ly\alpha}})
\end{align}
where $\lambda_{\mathrm{Ly\alpha}} = 1215.67$ \AA.
We assume the noise is uncorrelated at different wavelengths, which results in a diagonal noise matrix with $N_{ii}=\sigma_i^2$, where $\sigma_i$ is the continuum-normalized pipeline noise.

\subsubsection{Resolution matrix}
\label{subsubsec:resolution}
In our previous applications of QMLE, we approximated the spectrograph resolution effects by a continuous window function $W(v, v')$ such that the smoothed flux fluctuations $\delta_R$ were given by
\begin{equation}
    \delta_R(v) = \int \ddif v'\, W(v, v') \delta(v').
\end{equation}
Even though it is a valid and prevalent approximation, this formalism unfortunately fails to capture wavelength dependent resolution of the spectrographs.
However, for DESI the spectral extraction is built on the improved spectro-perfectionism algorithm \citep{boltonSpectroPerfectionismAlgorithmicFramework2010, guySpectroscopicDataProcessingPipeline2022}.
Spectro-perfectionism produces a resolution matrix $\mathbf{R}$ associated with each spectrum that is based on the spectrograph resolution as well as the noise properties of each spectrum, and captures the wavelength dependent resolution on the same discrete wavelength bins as the spectrum.
The observed signal becomes a matrix-vector multiplication: 
\begin{equation}
    \bm{\delta}_R = \mathbf{R} \bm{\delta}.
\end{equation}
This redefinition is natural to incorporate into the QMLE formalism. 
We achieve this by replacing the integral equations for the signal $\mathbf{S}$ and derivative matrices $\mathbf{Q}$ by the following expressions:
\begin{align}
    \mathbf{S}_R &= \langle \bm{\delta}_R \bm{\delta}_R^T \rangle 
    = \mathbf{R} \mathbf{S} \mathbf{R}^T \\
    \mathbf{Q}^\alpha_R &= \mathbf{R} \mathbf{Q}^\alpha \mathbf{R}^T,
\end{align}
where the subscript $R$ denotes the smoothed matrices, and matrices without a subscript are evaluated as integrals but now without a resolution window function.
\begin{equation}
    S_{ij}^{\mathrm{fid}} = \int_0^\infty \frac{dk}{\pi} \cos(k v_{ij}) P_{\mathrm{fid}}(k, z_{ij}),
\end{equation}
where $v_{ij}\equiv v_i - v_j$ and $1 + z_{ij} \equiv \sqrt{(1+z_i)(1+z_j)}$. 
The derivative matrix for redshift bin $m$ and wavenumber bin $n$ is
\begin{equation}
    Q_{ij}^{(mn)} = I_m(z_{ij}) \int_{k_n}^{k_{n+1}} \frac{dk}{\pi} \cos(kv_{ij}),
\end{equation}
where $I_m(z)$ is the interpolation kernel. This is 1 when $z=z_m$ and 0 when $z=z_{m\pm1}$. 

However, there are more subtleties regarding the resolution matrix.
First, these matrix multiplications require that all pixels are present, so we mark masked pixels with large noise estimates instead of eliminating them from the spectrum.
Second, the resolution matrix does not capture the resolution outside the spectral range (by construction).
This is a potential problem at the largest scales, so we implement an option in QMLE that pads the resolution matrix by mirroring its columns at the edges.
Third, both synthetic spectra and the actual DESI pipeline produce this matrix on the same wavelength grid as the spectrum with the same spacing.
This is natural in the spectro-perfectionism formalism in data, and we test its accuracy in Section~\ref{subsec:ccd-sim}; 
however, it yields an undercorrection at small scales in the mock analysis.
Our solution to this problem in mocks is to oversample every row of the resolution matrix \citep[Appendix D,][]{guySpectroscopicDataProcessingPipeline2022}.
One could model the resolution matrix at each row (i.e. wavelength) as a convolution of Gaussian and top-hat window functions, and fit for one or two free parameters for this model. One then evaluates each row of the oversampled resolution matrix using the best-fitting parameters at smaller wavelength steps. However, spectro-perfectionist resolution matrix carries negative elements and evidently does not follow this simple description.
Therefore, achieving a stable oversampling method requires a nuanced procedure.
We decided to use an unassuming description by interpolating the intermediate values. To correctly capture the rapid change in resolution matrix elements, we interpolate using their natural logarithms with a cubic spline. To obtain a valid natural logarithm, we shift every element to a small positive value in each row.
This small positive value is the smallest absolute value in that row (using an arbitrary number breaks down in subsequent steps).
We then apply a cubic spline to the natural logarithm of these elements,  oversample at a desired factor (usually three), and finally trace back these changes to obtain the new resolution matrix.

\subsubsection{Shifting Nyquist frequency on a linear wavelength grid}
Another update to QMLE concerns the Fisher matrix and DESI's wavelength binning.
The DESI pipeline extracts spectra on a linear wavelength grid of $\Delta \lambda=0.8$\,\AA, which results in an increasing Nyquist frequency with wavelength in velocity space $k_\mathrm{Ny}=\pi/dv$, where $dv=c\Delta\lambda/\lambda$.
In other words, we can measure higher $k$ modes at higher redshifts.
However, forcing the code to measure the same $k$ bins at lower redshifts results in numerically unstable Fisher matrix elements that could contaminate all scales when inverted.
Hence, these modes should be removed from the analysis.

We decide each spectral segment's Nyquist frequency using their median $dv$,
then set $k>k_\mathrm{Ny}/2$ modes in the Fisher matrix and the power spectrum to zero.
Since this procedure results in a "singular" matrix, we replace zeros on the diagonal with one while inverting.
Note this replacement does not contaminate lower $k$ modes, 
because it constitutes a block diagonal matrix.
This process stabilises the Fisher matrix.

\begin{figure*}
    \centering
    \includegraphics[width=0.8\linewidth]{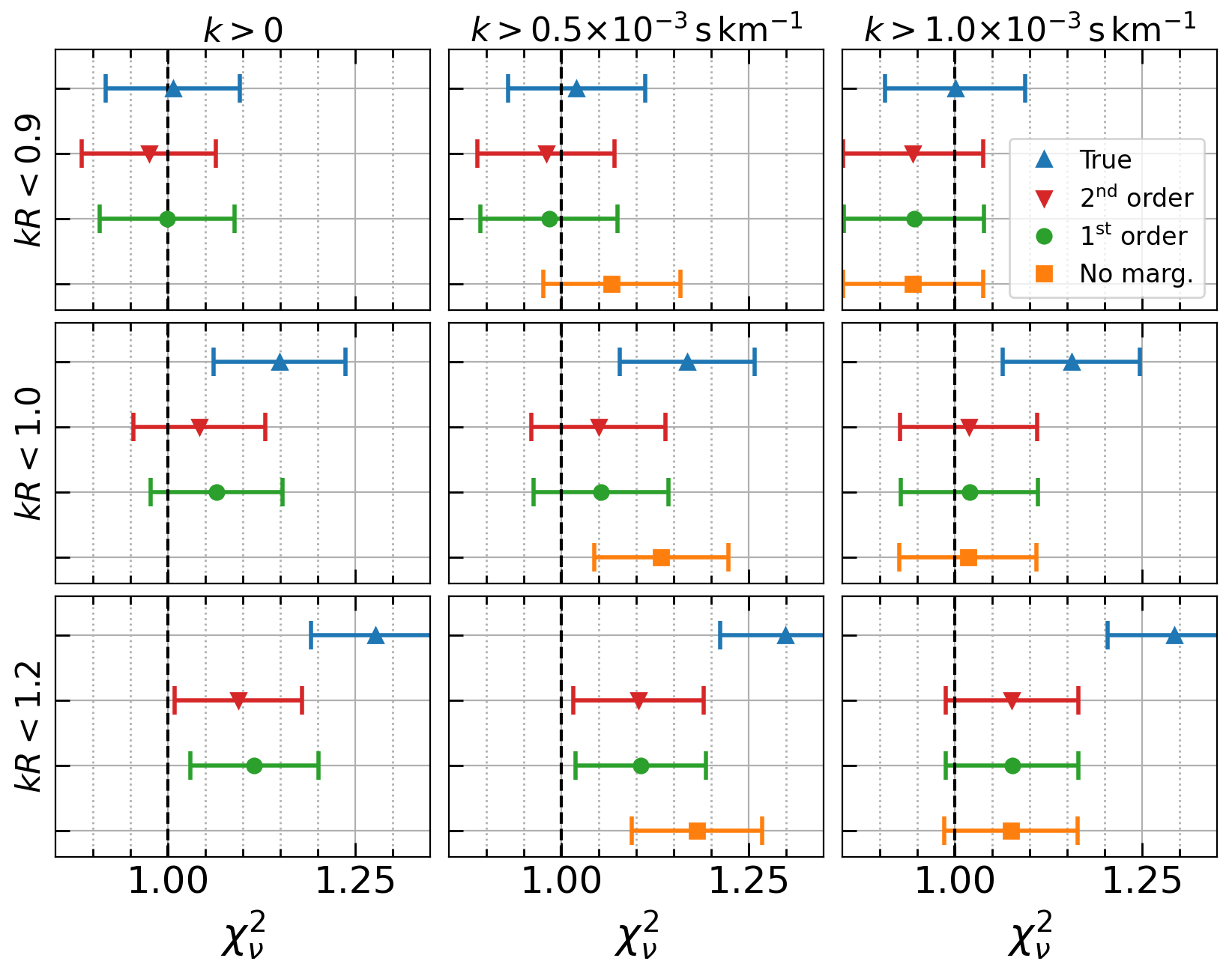}
    \caption{Reduced $\chi^2_\nu$ comparison for different $k$ cuts and continuum marginalisation polynomials on mocks.
    We find $\chi^2_\nu$ increases for all settings as we include higher $k$ values, which is unfortunate but expected since our correction to the \texttt{quickquasars} resolution matrix is not exact.
    The true continuum analysis results ({\it blue triangles}) stay within $1.5\sigma$ of $\chi^2_\nu=1$.
    Lower rows correspond to larger small-scale confidence regions.
    From left-most column to the right, we remove large-scale modes.
    When continuum errors are not marginalised ({\it orange squares}), throwing out these large-scale modes brings $\chi^2_\nu$ down to 1 within error bars.
    We also find that first ({\it green circles}) and second ({\it red triangle}) order marginalizations remove the contamination from continuum errors at all scales.
    }
    \label{fig:cont-marg-chi2}
\end{figure*}

\subsubsection{Nominal estimator settings}
Throughout this paper, we use 20 linear bins with $\Delta k_\mathrm{lin}=0.5\times10^{-3}$\skm and 13 log-linear bins with $\Delta k_\mathrm{log}=0.05$.
We use redshift bins of size $\Delta z=0.2$ from $z=2.0$ to $z=3.8$ included.
To reduce computation time and help continuum marginalisation, we split the spectra into two segments if they have more than 500 pixels, and we ignore segments having less than 20 remaining pixels.
We interpolate the signal and derivative matrices using 3601 points in velocity with 10\kms spacing and 400 points in redshift.

\subsection{Software}
Our quadratic estimator\footnote{\url{https://github.com/p-slash/lyspeq}} is written in \textsc{c++}.
It depends on \textsc{cblas} and \textsc{lapacke} routines for matrix/vector operations, \textsc{GSL}\footnote{\url{https://www.gnu.org/software/gsl}} for certain scientific calculations \citep{GSL}, \textsc{FFTW3}\footnote{\url{https://fftw.org}} for deconvolution when needed \citep{FFTW05}; and uses the Message Passing Interface (MPI) standard\footnote{\url{https://www.mpi-forum.org}}$^,$\footnote{\url{https://www.mpich.org}}$^,$\footnote{\url{https://www.open-mpi.org}} to parallelize tasks.
The DESI spectra are organized using HEALPix \citep{healpix} scheme on the sky.
We use the following commonly-used software in \textsc{python} analysis: \textsc{astropy}\footnote{\url{https://www.astropy.org}}
a community-developed core \textsc{python} package for Astronomy \citep{astropy:2013, astropy:2018, astropy:2022},
\textsc{numpy}\footnote{\url{https://numpy.org}}
an open source project aiming to enable numerical computing with \textsc{python} \citep{numpy},
\textsc{scipy}\footnote{\url{https://scipy.org}}
an open source project with algorithms for scientific computing,
\textsc{healpy} to interface with HEALPix in \textsc{python} \citep{healpy},
\textsc{numba}\footnote{\url{https://numba.pydata.org}}
an open source just-in-time (JIT) compiler that translates a subset of \textsc{python} and \textsc{numpy} code into fast machine code,
\textsc{mpi4py}\footnote{\url{https://mpi4py.readthedocs.io}}
which provides \textsc{python} bindings for the MPI standard \citep{mpi4py}.
Finally, we make plots using
\textsc{matplotlib}\footnote{\url{https://matplotlib.org}}
a comprehensive library for creating static, animated, and interactive visualizations in \textsc{python}
\citep{matplotlib}.

\section{Validation}
\label{sec:validation}
Synthetic data are crucial to verify that the measurements are unbiased, and the errors are correctly captured.
Our mock generation procedure consists of the generation of transmission files with forest fluctuations, diverse quasar spectra, and simulation of the DESI instrument.
The lognormal mock transmission files are generated using the procedure in \citet{karacayliOptimal1DLy2020}.
We generate them on a linear wavelength grid of 0.2\,\AA\ spacing without any resolution and noise effects.

We develop two methods to simulate and validate the DESI analysis pipeline. 
The first set of mocks is produced using \texttt{quickquasars}, which is part of the \textsc{desisim} package\footnote{\url{https://github.com/desihub/desisim}} and uses \textsc{specsim}\footnote{\url{https://github.com/desihub/specsim}} \citep{kirkbySpecsim2021} for quick simulations of fiber spectrograph response (see \citet{herrera-alcantarDESIQuickquasars2023} for a detailed description of \texttt{quickquasars} mocks).
This program generates random quasar continua, simulates sky and instrumental noise, and incorporates wavelength-dependent camera resolution, but does not validate the computationally expensive spectral extraction.
Hence, we cannot validate the spectro-perfectionist resolution matrix with these mocks.
In order to apply and validate the spectro-perfectionism algorithm in the \lya forest,
we create a second set of mocks called "CCD image simulations" that project mock quasar spectra onto two-dimensional images that simulate DESI raw data at the CCD pixel level with the \textsc{desisim} package.
These CCD image simulations are then processed in a similar manner to actual data with the algorithms that comprise the DESI spectroscopic reduction pipeline \citep{guySpectroscopicDataProcessingPipeline2022}. This approach is more computationally expensive than one-dimensional mocks, so we only employ it on a smaller number of mock spectra.

\subsection{Quickquasars mocks}
\label{subsec:quickquasars}
For these mocks, the quasar diversity, DESI instrument and the sky are simulated through a program called \texttt{quickquasars} in the \textsc{desisim} package.
This program randomly generates quasar continua from a broken power law with emission lines, convolves with the wavelength-dependent camera resolution for each arm, adds noise for a given exposure time and observation program, and finally resamples onto the output DESI wavelength grid of $\Delta\lambda_\mathrm{DESI}=0.8$\,\AA\ per pixel.
We smooth out the source contribution to noise with a Gaussian kernel of $\sigma=10$\,\AA\ to imitate the DESI pipeline \citep{guySpectroscopicDataProcessingPipeline2022}.

All unique targets that are identified as quasars are simulated in our mocks.
However, in real data analysis, we remove certain surveys, programs and low SNR targets.
We generate the transmission files with the exact redshift distribution of DESI quasars in our sample and assume a constant $4000\,s$ exposure time for all spectra.

As extensive and realistic as \texttt{quickquasars} is, it does not fully reproduce the spectral extraction pipeline output since it does not generate 2D CCD images. As an important consequence, the output resolution matrix does not follow the spectro-perfectionism formalism and instead it is approximated as a box-car average over rows and columns of the finely sampled camera resolution matrix.
Unfortunately, this approximation is not correct as it smoothes the resolution matrix twice, once over rows and once over columns.
To correct that implementation, we deconvolve a top-hat window function and oversample each row of this matrix by a factor of three in the power spectrum estimation. This yields adequately unbiased power spectrum results but is not a precise enough solution to strongly rely on $\chi^2$ criteria.
We also perform CCD image simulations to understand the behaviour of the resolution matrix in data.

As noted, we generate a mock spectrum for each unique target in our sample, which yields 92\,780 quasars in our final sample.
We define the forest to be between 1050--1180~\AA\ in the quasar rest frame, use the analytically calculated true power spectrum as our fiducial and perform a single iteration using the QMLE \citep{karacayliOptimal1DLy2020}.
We define our small-scale confidence range with respect to effective velocity spacing $R_z=c\Delta\lambda_\mathrm{DESI}/(1+z)\lambda_\mathrm{Ly\alpha}$ of each redshift bin, where $\Delta\lambda_\mathrm{DESI}=0.8$\,\AA.

\subsubsection{True continuum, no systematics}
We start validating our analysis without any continuum fitting complications or other systematics. 
We obtain flux transmission fluctuations $\delta_F$ using the true continuum (which is provided by \texttt{quickquasars}) and true mean transmission. We estimate the mean transmission from pure transmission files, confirm that this estimate is correct using the analytical mean transmission expression \citep{karacayliOptimal1DLy2020}, and use the analytical expression to remove the measurement noise.

We find that the estimated power spectrum agrees with the true underlying power albeit the problems at small scales due to inaccuracy in the resolution as mentioned above.
We calculate the reduced chi-square $\chi^2_\nu = \chi^2 / \nu$, where the number of degrees of freedom $\nu$ is equal to the number of $P(k, z)$ points in the range of concern, and $\chi^2 = (\bm{P} - \bm{P}_\mathrm{true})^T \mathbf{C}^{-1}(\bm{P} - \bm{P}_\mathrm{true})$ where $\bm P$ is the measurement, $\bm P_\mathrm{true}$ is the underlying true power spectrum and $\mathbf{C}$ is the covariance matrix from QMLE.
Figure~\ref{fig:cont-marg-chi2} shows the reduced $\chi^2_\nu$ from the true continuum analysis results in blue triangles. $\chi^2_\nu$ values increase as we include higher $k$ values (going from the top to the bottom row), which is unfortunate but expected since our correction to the \texttt{quickquasars} resolution matrix is not exact. The $k R < 0.9$ range is firmly validated with $\chi^2_\nu \approx 1$. The $k R < 1$ range deteriorates the agreement between power spectra by $1.5\sigma$, and finally, the agreement breaks down in the $kR < 1.2$ range.

\begin{figure}
    \centering
    \includegraphics[width=\columnwidth]{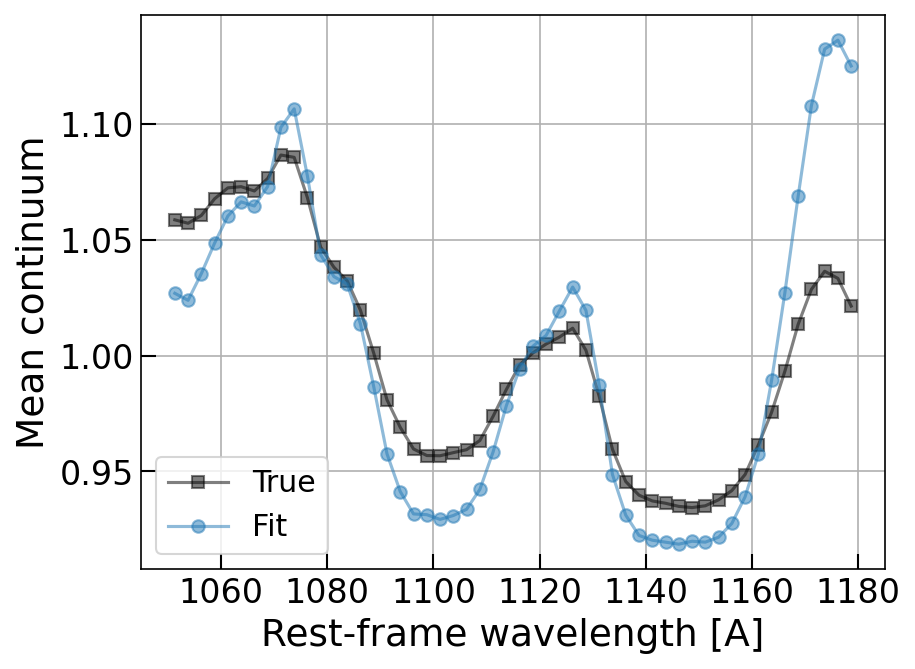}\\
    \includegraphics[width=\columnwidth]{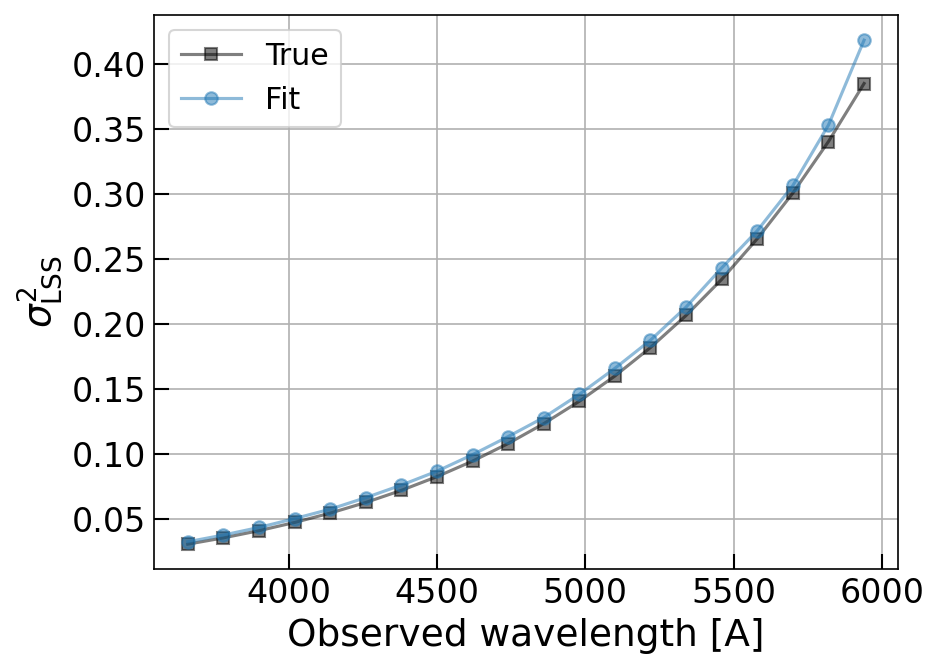}
    \caption{({\it Top}) Mean global continuum from the true continuum analysis vs.\ from continuum fitting on mocks.
    The continuum fitting accentuates features in the mean continuum.
    ({\it Bottom}) $\sigma^2_\mathrm{LSS}$ values from the true continuum analysis vs.\ continuum fitting.
    Our fitting algorithm finds the correct $\sigma^2_\mathrm{LSS}$ values.}
    \label{fig:meancont_varlss_mock}
\end{figure}

\subsubsection{Continuum fitting, no systematics}
We now turn to validating our continuum fitting procedure, since we do not have access to the true quasar continua.
The \texttt{quickquasars} code generates quasar continua with broken power laws and emission lines, so our continuum fitting model with a single global mean and two diversity parameters is not exact.
Therefore, our mock continuum is not tailored toward our fitting model, and the test results we present here also capture model mismatches.

There are $92,780$ quasars in our mock data set. We find that fitting for $\sigma_\mathrm{LSS}$ is not valid for observed wavelength $\lambda>6000\,$\AA\ due to the small number of high-redshift quasars with forest data at these wavelengths (only $883$), so we limit our continuum fitting region to $3600-6000$\,\AA. This sets $z_\mathrm{Ly\alpha}=3.8$ as our largest redshift bin.
We measure the global mean continuum $\overline{C}(\lambda_\mathrm{RF})$ in 2.5\AA\, steps.
We fix $\eta=1$, and measure $\sigma^2_\mathrm{LSS}$ in 20 wavelength bins in the observed frame in equation~(\ref{eq:sigma2_cfit}).
We do not apply a SNR cut in order to keep all spectra and perform five iterations.

In Figure~\ref{fig:meancont_varlss_mock}, we compare the mean continua from the true continuum analysis to the one from continuum fitting.
Continuum fitting accentuates peaks and valleys in the mean continuum compared to when the true continuum is known.
These deviations are interesting and merit further investigation, but our main objective is to obtain unbiased \poned results.
As we discuss below, these deviations do not impede that objective.
The bottom panel of Figure~\ref{fig:meancont_varlss_mock} shows $\sigma^2_\mathrm{LSS}$ estimated from the true and continuum fitting analyses.
We find fitting the variance leads to the correct $\sigma^2_\mathrm{LSS}$ values.
We note that $\sigma^2_\mathrm{LSS}$ is not only a function of \poned, but also depends on spectrograph resolution and wavelength spacing.

\begin{figure}
    \centering
    \includegraphics[width=\columnwidth]{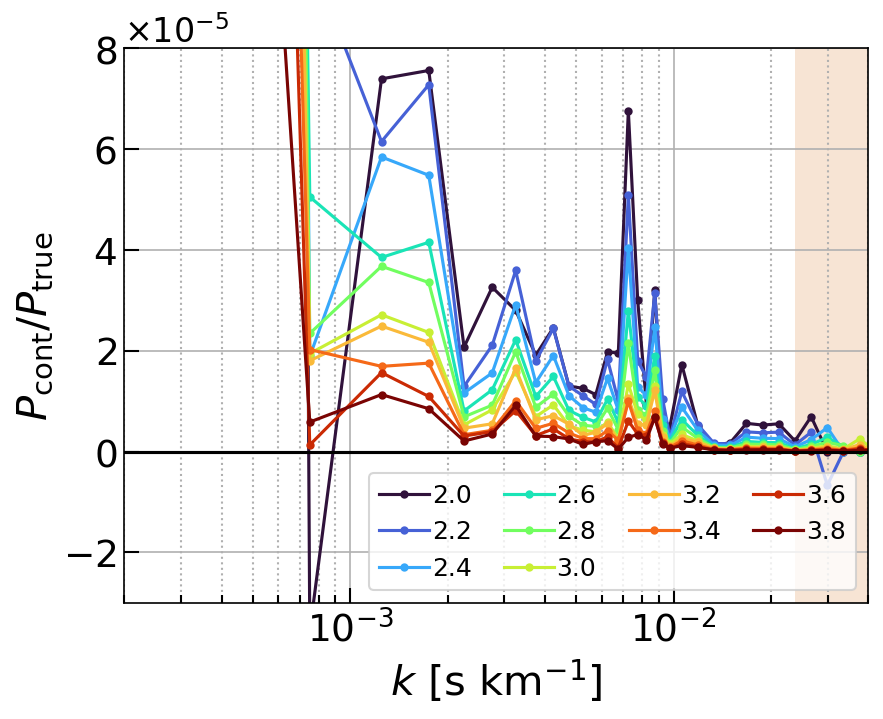}
    \caption{Power spectrum of the remaining continuum errors after marginalization, divided by the true underlying mock \lya power spectrum.
    We find the remaining continuum errors are a factor of $10^{-5}$ smaller than the signal at most scales and redshifts.
    The maximum is $10^{-3}$ at the largest scales.}
    \label{fig:mock-continuum-err-power}
\end{figure}

We investigate reduced $\chi^2_\nu$ values for various settings to judge the accuracy of the \poned estimate.
In addition to the true continuum results, Figure~\ref{fig:cont-marg-chi2} shows results for no continuum marginalization (orange squares), first order $\ln \lambda$ polynomial (green circles), and second order polynomical (red triangles).
$\chi^2_\nu$ from no the marginalization case is not visible in the left-most column, but produces reasonable values when large-scale modes are removed in the middle and right columns.
This result illustrates the importance of continuum marginalization, especially in that we can retain even the largest scales.
We note that this analysis does not account for metals or DLA systematics, which dominate at these scales.
Furthermore, we estimate the power spectrum produced by the remaining continuum errors by calculating $\delta_C \equiv C_\mathrm{est}/C_\mathrm{true}-1$ and running it through QMLE.
We do not subtract noise and fiducial terms in this case, but keep everything else the same.
We find this continuum error power spectrum is a factor of $10^{-5}$ smaller than the signal at most scales and redshifts as shown in Figure~\ref{fig:mock-continuum-err-power}.
With these results, we consider our continuum fitting and marginalization validated for this work.
In future work, we will test our analysis pipeline on multiple (ideally 100) realizations and directly study the $\chi^2$ distribution.

\subsubsection{Masking high-column density systems}
We finally tested masking the high-column density (HCD) systems both in continuum fitting and in the \poned estimate.
We generate mocks with randomly placed HCDs and build a truth catalogue using their redshifts and column densities.
There are 17\,273 HCDs with $N_\mathrm{\ion{H}{I}}>19.5$ on 15\,097 sightlines, which corresponds to 16.2\% of all sightlines.

We calculate a DLA transmission profile based on the column density for each system. The damping wings extend to large wavelength separations from the central wavelength, such that an aggressive masking strategy would remove many data points. Therefore, we mask pixels where the model profile is below a transmission threshold and correct the damping wings at larger transmission values based on the same model profile. A higher transmission threshold results in smaller corrections, but eliminates more data points. We tested two thresholds: a nominal threshold with the DLA absorption greater than 20\% and a conservative cut of greater than 10\%. We find both options yield similar $\chi^2_\nu\sim 1$ within error bars.

\begin{figure}
    \centering
    \includegraphics[width=\columnwidth]{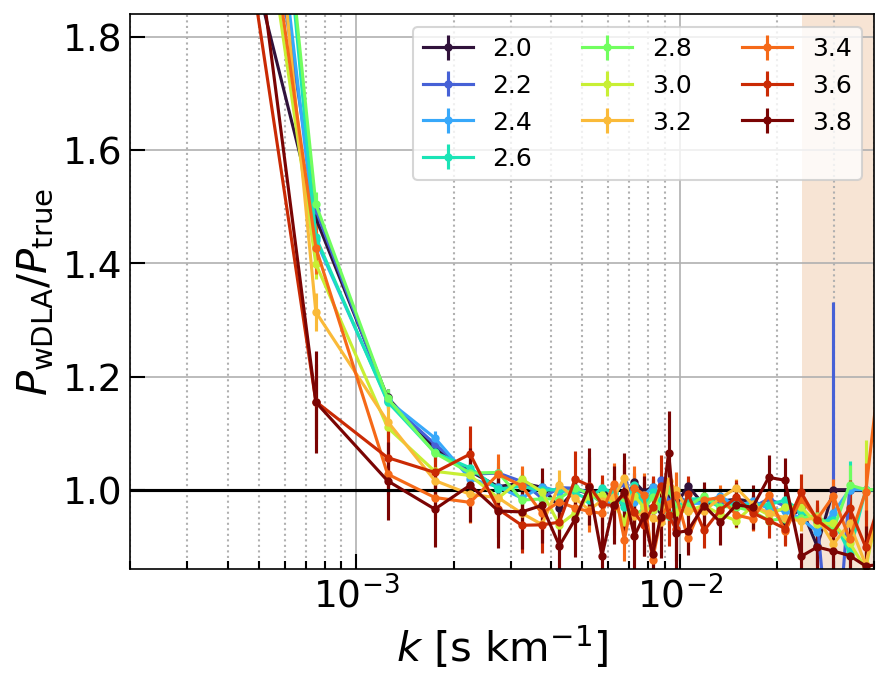}
    \caption{Power spectrum when DLAs and sub-DLAs are not masked divided by the true underlying mock power spectrum.
    These systems add power to large scales.
    }
    \label{fig:mock-nodla-wrt-true-ratio}
\end{figure}

When unmasked, these systems add power to the large scales, as shown in Figure~\ref{fig:mock-nodla-wrt-true-ratio}. Furthermore, this extra power depends on whether these systems are correlated with the underlying matter field \citep{mcdonaldPhysicalDampingWings2005}, and it has different amplitudes and shapes for different column densities \citep{rogersEffectsOfHcd2018}. Accurate simulation of these systems embedded in the \lya\ forest remains challenging, yet they cannot be completely removed from the measurement either. For instance, the catalogue produced by the DLA finder is approximately 90\% efficient and pure \citep{wangDeepLearningDESIDLA2022}.
As we discuss in Section~\ref{subsec:systematics}, we report the systematics associated with the DLA finder inefficiency based on a simple scaling of this ratio. However, the effect of uncorrected and undetected DLAs still needs to be modelled and marginalized over in cosmological inferences \citep{palanque-delabrouilleOnedimensionalLyalphaForest2013, chabanierOnedimensionalPowerSpectrum2019}.

\subsection{CCD image simulations}
\label{subsec:ccd-sim}

We validated the resolution matrix implementation in the DESI pipeline and the QMLE with CCD image simulations. The image simulations draw on extensive infrastructure in the \textsc{desisim} package that was built by the DESI team to develop and validate the spectroscopic data processing pipeline \citep{guySpectroscopicDataProcessingPipeline2022} in advance of first light. This package produces realistic two-dimensional spectroscopic image simulations that include the bias, readnoise, and gain for each amplifier for each of the 30 CCDs, models the throughput based on engineering data for each channel of each spectrograph, the model PSF and trace of each fiber as a function of wavelength and position on the detector based on the Zemax optical design models, sky emission, and applies noise appropriate for the flux of each object at the desired exposure time.

The typical input to the simulation code is a library of the model spectra for a single DESI observation and a file that describes how these objects are distributed into the fibers. This mapping of targets to fiber positions for one observation corresponds to a single DESI tile. Normally, a tile would include all of the DESI target classes, along with a selection of flux calibration stars and designated `empty' fibers that are used to measure the night sky spectrum. Since we are just interested in validating the performance with the \lya forest and the time to construct one simulation tile of 30 CCD images is not insignificant, each of our tile designs has only quasars, standard stars, and sky fibers. There are approximately 4500 quasars in each tile, and the quasars are all at $2.6 < z < 3.6$ so we observe the entire forest region for each quasar in the blue channel of the spectrographs. We generate ten tiles, or approximately 45,000 total \lya forest spectra, with apparent magnitudes representative of DESI quasar targets and noise that is representative of a single, 1000\,s exposure in nominal dark conditions. We also generate a mock set of arc calibration lamp exposures, which are used by the pipeline to measure the PSF and for wavelength calibration.

We processed this simulated dataset with the DESI pipeline, which is described in detail by \citet{guySpectroscopicDataProcessingPipeline2022}. Briefly, the pipeline pre-processes the raw images to remove the bias level, fits the arc calibration lamps to measure the nightly PSF, spectral trace, and wavelength calibration, refits the nightly PSF to the PSF of each exposure, extracts the spectra, applies a flat field, subtracts the sky, fits the flux calibration stars, applies the flux calibration, and measures the redshift and identifies the best spectral type for each target. The pipeline fits the PSF with an empirical model that consists of a linear combination of Gauss-Hermite functions. The pipeline model is an excellent but not exact match to true PSF, which is well-approximated by the (non-parametric) PSF predicted by the optical design, and we generated the simulated arc and observation files with the optical model PSF to include any systematic error associated with the pipeline's PSF model fit in our analysis. To isolate the impact of fitting the PSF model from the remainder of the spectral extraction and calibration steps, we also processed the simulations with the correct PSF model.

\begin{figure}
    \centering
    \includegraphics[width=\columnwidth]{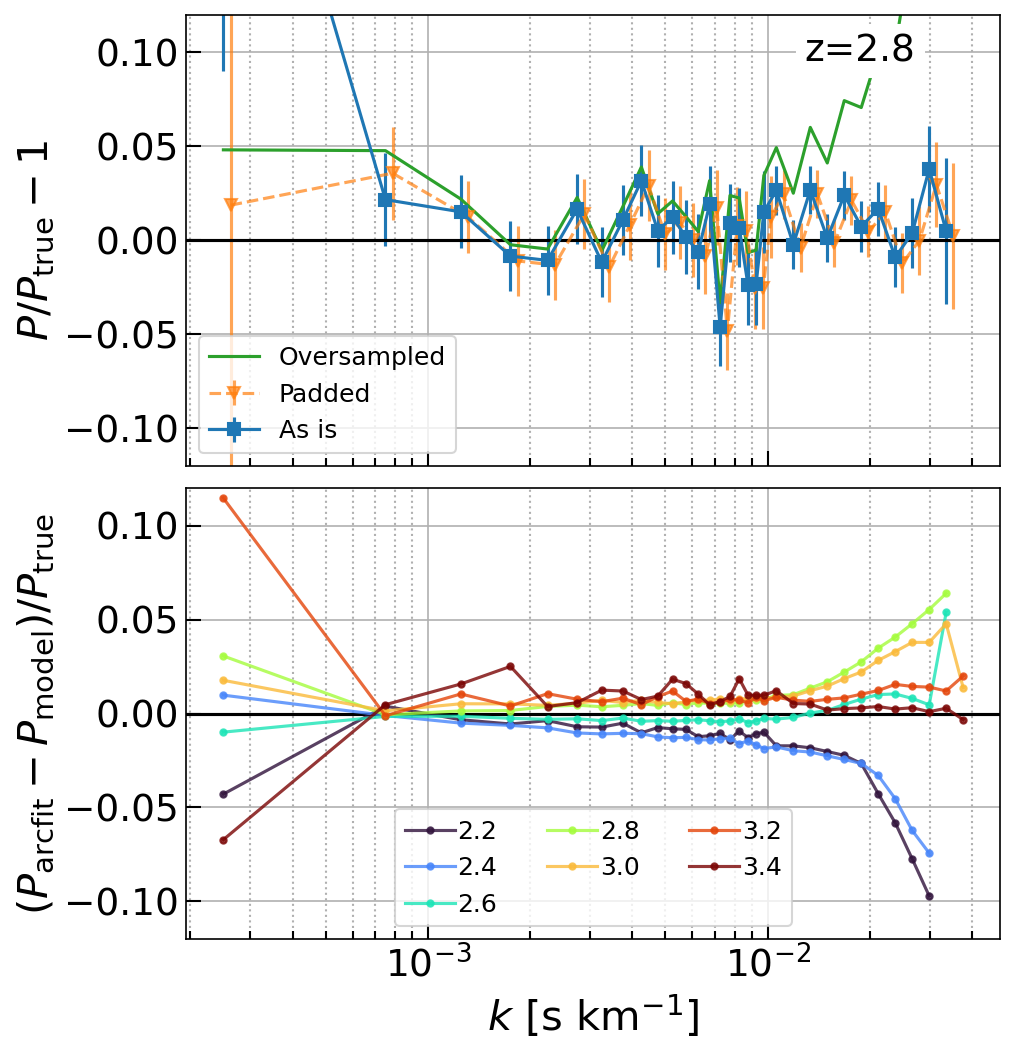}
    \caption{({\it Top}) Power spectrum estimates for various resolution matrix treatments at $z=2.8$ based on CCD image simulations.
    The default pipeline output ({\it blue squares}) performs best, whereas oversampling ({\it green line}) deviates from the truth.
    Padding the edges of the resolution matrix ({\it orange triangles}) improves the agreement at the largest scales, but those modes are lost to continuum errors in any case.
    ({\it Bottom}) The difference between power spectrum estimates using the fitted PSF and the true model PSF, divided by the true underlying signal.
    The fitted PSF introduces a wavelength dependent resolution error.
    }
    \label{fig:ccd-image-reso}
\end{figure}

Figure~\ref{fig:ccd-image-reso} ({\it top}) shows the results for the case where the pipeline starts with the correct PSF model. This panel shows the ratio of the measured power spectrum $P$ to the input power spectrum $P_\mathrm{true}$ based on the resolution matrix provided ``as is'' by the pipeline (blue squares) for a subset of the quasars at $z = 2.8$. The ratio is consistent with unity over the entire range in $k$, with the exception of the largest scales (smallest $k$), which are in any case not usable due to continuum fitting errors. We also explored oversampling the resolution matrix to better match the input model (see Section~\ref{subsubsec:resolution} for details), and padding the resolution matrix to remove edge effects. None of these modifications is superior to using the resolution matrix provided by the pipeline.

We also analyzed the performance of the pipeline with an empirical PSF measured from the arc calibration lamps, rather than with the input PSF model as in the previous case. The bottom panel of Figure~\ref{fig:ccd-image-reso} shows the difference in the 1D power spectrum measurement starting with the arc calibration lamps $P_\mathrm{arcfit}$ and starting with the correct model PSF $P_\mathrm{model}$ for seven redshift bins from 2.2 to 3.4. The fractional difference relative to the true power spectrum is at the level of 1--2\% up to $k=0.01$\skm after which the errors grow exponentially as expected. This behaviour is consistent with 1\% precision on the resolution itself. This is our best estimate of the systematic error contribution of the resolution matrix to the measurement.

A potential limitation of these simulations is that real DESI observations have mostly galaxies, rather than quasars. While our simulations have typical numbers of standard stars and sky fibers, they have no galaxies and consequently do not simulate potential cross-talk between galaxy and quasar targets. \citet{guySpectroscopicDataProcessingPipeline2022} carefully studied cross talk between adjacent fibers and found that this is minimal, even for bright calibration lines, so we do not anticipate this will be important for much fainter the continuum and emission lines present in galaxies and quasars. 

\begin{figure*}
    \centering
    \includegraphics[width=0.95\linewidth]{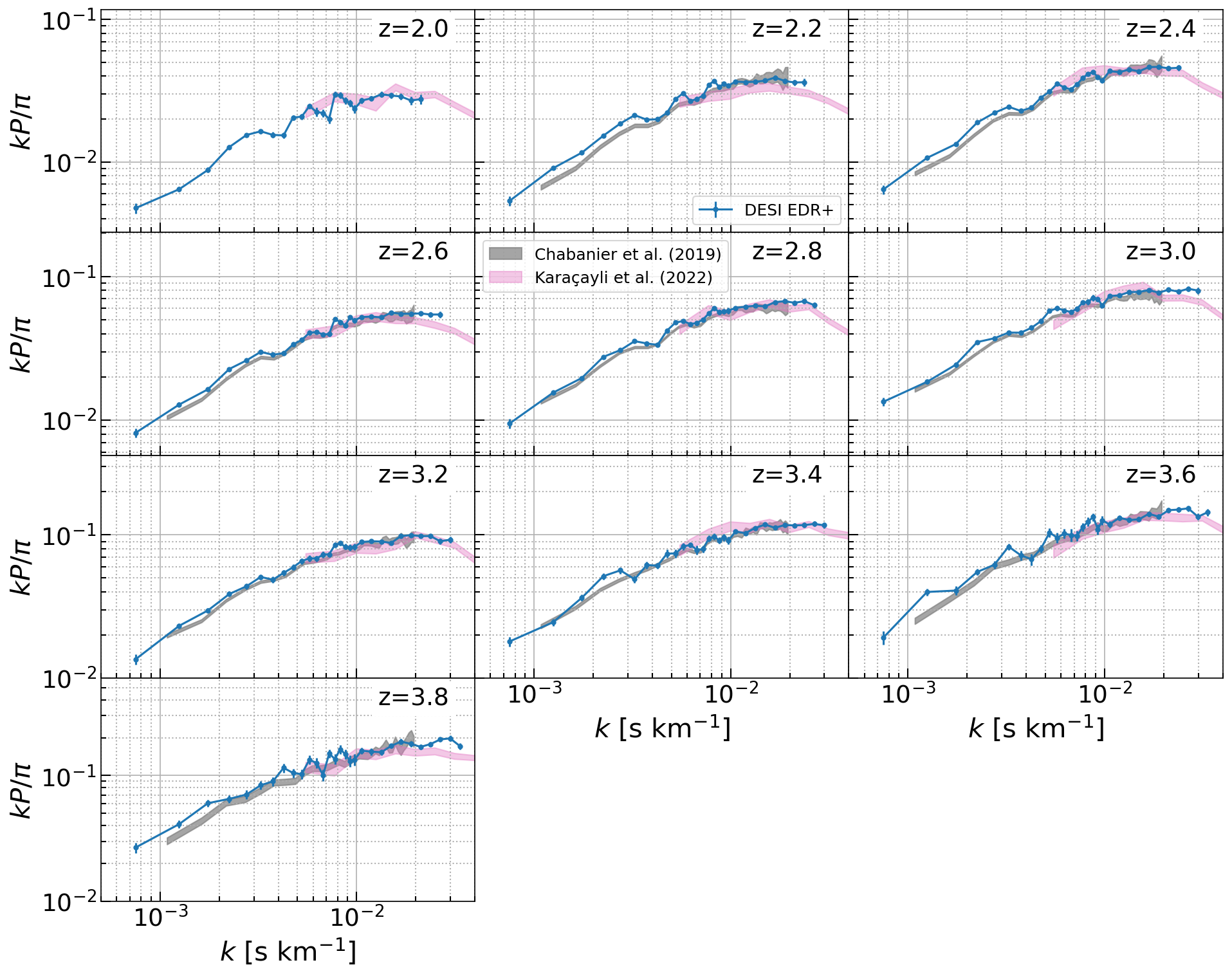}
    \caption{Final \lya forest \poned results.
    We remove BAL quasars from our sample, mask DLAs and major sky lines, and correct for pipeline noise and flux miscalibrations.
    Metal power is subtracted using side bands as described in Section~\ref{subsec:sidebands}.
    Error bars are from 200\,000 bootstrap realizations of 256 subsamples.
    Our \poned results are slightly larger than eBOSS measurements \citep{chabanierOnedimensionalPowerSpectrum2019}, which is most visible at $z=2.2$ and $2.4$ bins.
    }
    \label{fig:p1d-lya-sb1subt-grid}
\end{figure*}

\section{Results from data}
\label{sec:results}
After our comprehensive validation tests on the synthetic spectra, we now analyze DESI early data.
The \lya forest is measured in the $1050-1180$\,\AA\ rest-frame region of each quasar; and the global mean continuum is calculated using $\Delta \lambda_\mathrm{RF}=0.8$\,\AA\ coarse rest-frame binning pixels in this range.
We limit our analysis to the observed wavelength range of $3600-6000$\,\AA.
We also remove forests with mean SNR less than 0.25 in the forest region in order to minimize possible impurities in the quasar catalogue, where $\mathrm{SNR}\equiv F/\sigma_\mathrm{pipe}$\footnote{This definition of SNR is not ideal since it discriminates against high redshift forests which intrinsically have lower mean flux, and therefore lower SNR.
We will investigate improvements in the SNR definition in future work.}.
The locations of sky lines are naturally down-weighted by the pipeline, but we mask certain particularly strong lines because they are difficult to reliably model\footnote{\url{https://github.com/corentinravoux/p1desi/blob/main/etc/skylines/list_mask_p1d_DESI_EDR.txt}}.
DLAs are difficult to simulate accurately and complicate cosmological inference from $P_{\mathrm{1D}}$.
We thus mask DLAs at the wavelengths of their strongest absorption $F<0.8$ and correct for the damping wings (due to both \lya and Ly~$\beta$ transitions) above this threshold.
Furthermore, BAL features can contaminate the forest, and hence these features are also masked.
We do not estimate the pipeline noise calibration errors simultaneously (i.e. we fix $\eta=1$ in eqn.~\ref{eq:sigma2_cfit}) and only measure $\sigma^2_\mathrm{LSS}$ in 20 observed frame wavelength bins.
As we show below, the noise and flux reported by the pipeline have calibration errors, but due to heavy absorption and correlations between pixels, the \lya forest region is not stable to calibrate for these errors.
Instead, we carry out a meticulous study of statistics in the side band regions to calibrate our final reduction.
We limit continuum fitting to five iterations where we update the global mean continuum and $\sigma^2_\mathrm{LSS}$.

We estimate \poned using the following fiducial power parameters in equation~(\ref{eq:pd13_fitting_fn}): $A=0.066, n=-2.685, \alpha=-0.22, B=3.59, \beta=-0.16$ and $k_1=0.053$\,s\,km$^{-1}$.
We neither oversample nor pad the resolution matrix, and use it as provided by the pipeline.
We perform a single iteration to measure the power spectrum.
Further iterations mostly refine Fisher matrix estimates \citep{karacayliOptimal1DLy2020}, which we replace with a regularised bootstrap estimate as described below.
Even though we keep BAL quasars with masked features while fitting the continuum, since \citet{ennesserMitigationBALquasars2022} showed masking BAL lines yields an uncontaminated estimate of the mean continuum, we remove them from the \poned estimation to be conservative in our approach\footnote{Our preliminary comparison showed some deviation in \poned between two samples that we will explore in a future study.}.

Wavelengths larger than the \lya line in the spectrum are free from neutral hydrogen absorption, so they can be used to statistically estimate metal contamination and other systematics \citep{mcdonaldLyUpalphaForest2006, palanque-delabrouilleOnedimensionalLyalphaForest2013, chabanierOnedimensionalPowerSpectrum2019}.
The regions between strong emission lines at these wavelengths are called the "side bands" (SB).
Table~\ref{tab:qsonumbers} lists the wavelength ranges for the side bands and the number of quasars in each region.
Subtracting the SB 1 power spectrum statistically removes all power due to metals with $\lambda_\mathrm{RF}>1380$~\AA, but some metal contamination remains.
For example, the \ion{Si}{iii}-\lya cross-correlation imprints oscillatory features \citep{mcdonaldLyUpalphaForest2006, palanque-delabrouilleOnedimensionalLyalphaForest2013}.
We provide the details regarding the metal power estimations and other studies on systematics in the following subsections.

\begin{table}
    \centering
    \begin{tabular}{l|c|c}
         & Wavelength range [\AA] & \# All quasars \\
        \hline
        \lya & 1050--1180 & 54\,600 \\
        SB~1 & 1268--1380 & 115\,086 \\
        SB~2 & 1409--1523 & 153\,326 \\
        SB~3 & 1600--1800 & 194\,666
    \end{tabular}
    \caption{Rest-frame wavelength ranges and number of quasars in DESI early data.
    Quasars with BAL features are ignored.
    }
    \label{tab:qsonumbers}
\end{table}

\begin{figure}
    \centering
    \includegraphics[width=\columnwidth]{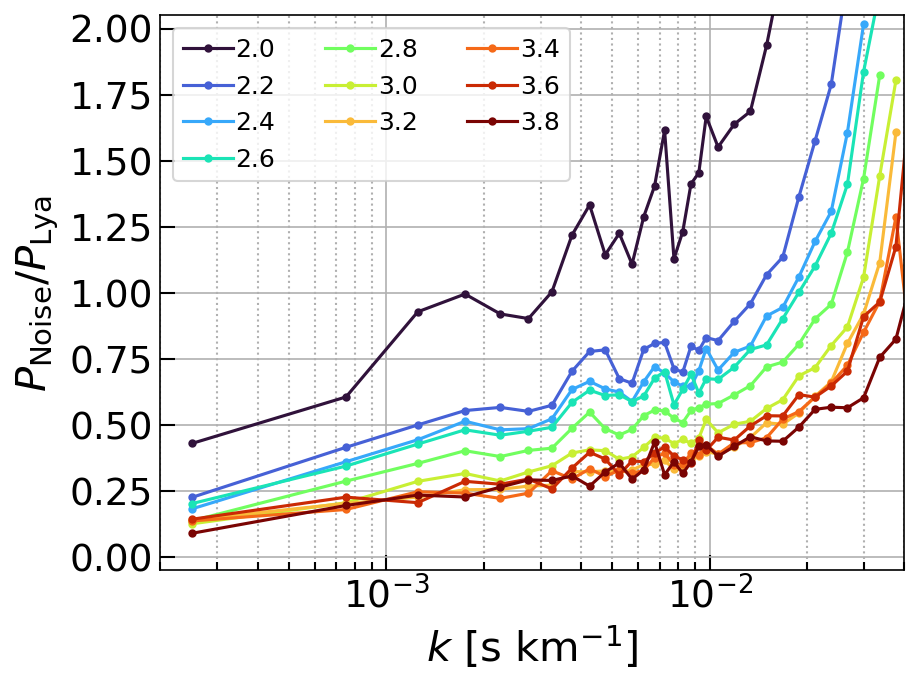}
    \caption{Ratio of noise power to noise-subtracted \lya power from data.
    The noise power spectrum is not negligible even at large scales.
    Results at $z=2.0$ are most sensitive to the noise power estimates. The features in this ratio mostly come from the inverse of \poned.}
    \label{fig:pnoise_plya_ratio}
\end{figure}

\begin{figure}
    \centering
    \includegraphics[width=\columnwidth]{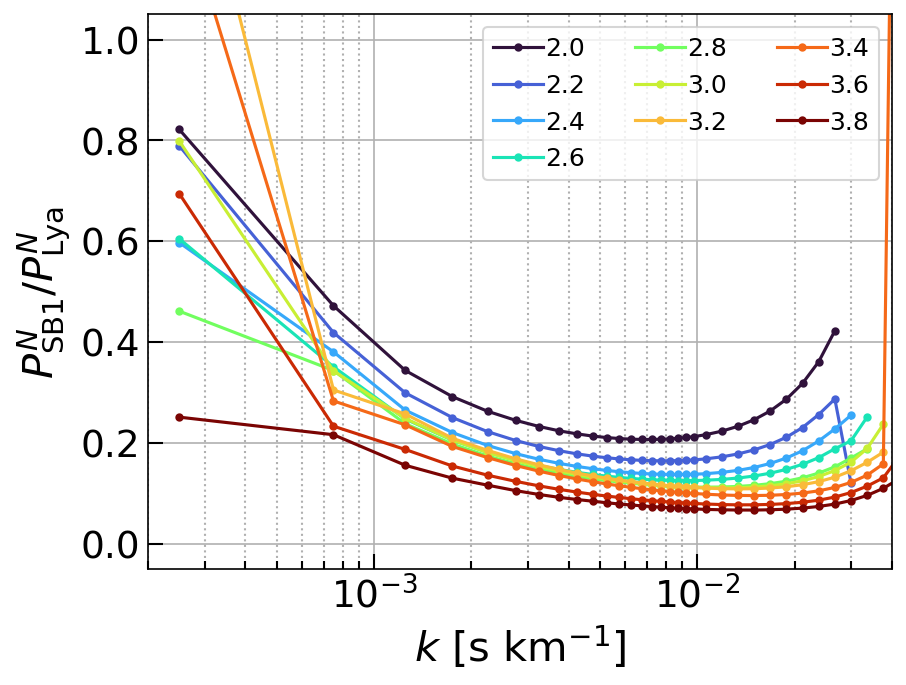}
    \caption{Ratio of noise power in SB 1 to \lya region from data.
    It is redshift and scale dependent.
    Noise power is smaller in side band regions, which means pipeline noise miscalibrations cannot be fully removed by side band subtraction.
    }
    \label{fig:pnoise_sb1_lya_ratio}
\end{figure}

The results after we subtract the metal power are shown in Figure~\ref{fig:p1d-lya-sb1subt-grid}.
These \poned results are $5-15\%$ larger than the eBOSS measurements \citep{chabanierOnedimensionalPowerSpectrum2019}, which corresponds to $1.5-3\sigma$ tension. This tension is most visible in the $z=2.2$ and $2.4$ bins, but is present in all redshift bins.
We present possible explanations for the origins of this discrepancy in Section~\ref{sec:discussion}.
Furthermore, accurate noise estimates are crucial to our final \poned results.
Figure~\ref{fig:pnoise_plya_ratio} shows the ratio of the noise power spectrum (equation~(\ref{eq:noise_bn})) to the noise-subtracted \lya power spectrum, which ranges between 25--100\% and is larger at higher $k$ values. The features in this figure can mostly be attributed to the inverse of \poned, but to be exact, QMLE's noise power spectrum is an inverse covariance weighted average and therefore manifests features based on the fiducial \lya power spectrum, continuum marginalization, and characteristics of pipeline noise. We stress that this noise power strictly comes from the randomness of observed flux values, and is not related to metal absorption, continuum fitting errors, DLA masking, or noise correlations between pixels.
Even small miscalibrations can directly propagate to our final \poned estimates.
The SB power subtraction can balance out some miscalibration, but not all of it.
We find that the SB noise power estimate is $10-25$\% of the \lya region as shown in Figure~\ref{fig:pnoise_sb1_lya_ratio}.
Fortunately, the pipeline noise estimates are accurate at the percent level and the remaining errors can be corrected by investigating the side bands.

\subsection{Bootstrap error estimates}
The Fisher matrix given by the QMLE assumes Gaussianity, and hence may not be representative of the statistical errors in the data either due to non-linearities in the \lya forest at small scales or other effects in quasar selection, DLA masking etc.
For that reason, we calculate the bootstrap covariance matrix for a more reliable error estimate as follows.
First, we save QMLE's power spectrum and Fisher matrix estimates in 256 sub-samples\footnote{Number of sub-samples is based on the MPI tasks used, which is a current code limitation.}.
We then estimate the bootstrap covariance using these sub-samples over $200\,000$ realisations.
As we noted in \citet{karacayliOptimal1DLy2022}, the bootstrap covariance is noisy (especially off-diagonal terms) and needs regularisation. We take advantage of the sparsity pattern of the covariance matrix \citep{padmanabhanSparseMatrices2016}, and regularise the bootstrap covariance as follows:
\begin{enumerate}
    \item We apply a sparsity pattern on the bootstrap covariance matrix using the Gaussian covariance matrix from QMLE such that $\left|r_{ij}^{\mathrm{QMLE}}\right|>r_\mathrm{min}$ and $r_{ij}\equiv C_{ij}/\sqrt{C_{ii}C_{jj}}$.
    \item We find the eigenvalues $\lambda_i$ and eigenvectors $\bm e_i$ of this sparse bootstrap covariance matrix.
    \item We calculate the precision of these eigenvectors under Gaussianity: $\lambda_i^{\mathrm{QMLE}} = \bm e_i^T \mathbf{C}^{\mathrm{QMLE}} \bm e_i$. This is the theoretical  minimum for the covariance.
    \item We replace $\lambda_i\rightarrow \mathrm{max}(\lambda_i, \lambda_i^{\mathrm{QMLE}})$ \citep{mcdonaldLyUpalphaForest2006}.
\end{enumerate}
We repeat these steps until convergence or for a maximum of 500 iterations.
We choose $r_\mathrm{min} = 0.01$ for \lya and $r_\mathrm{min} = 0.001$ for SB 1, because SB 1 values are more strongly correlated.

\begin{figure}
    \centering
    \includegraphics[width=\columnwidth]{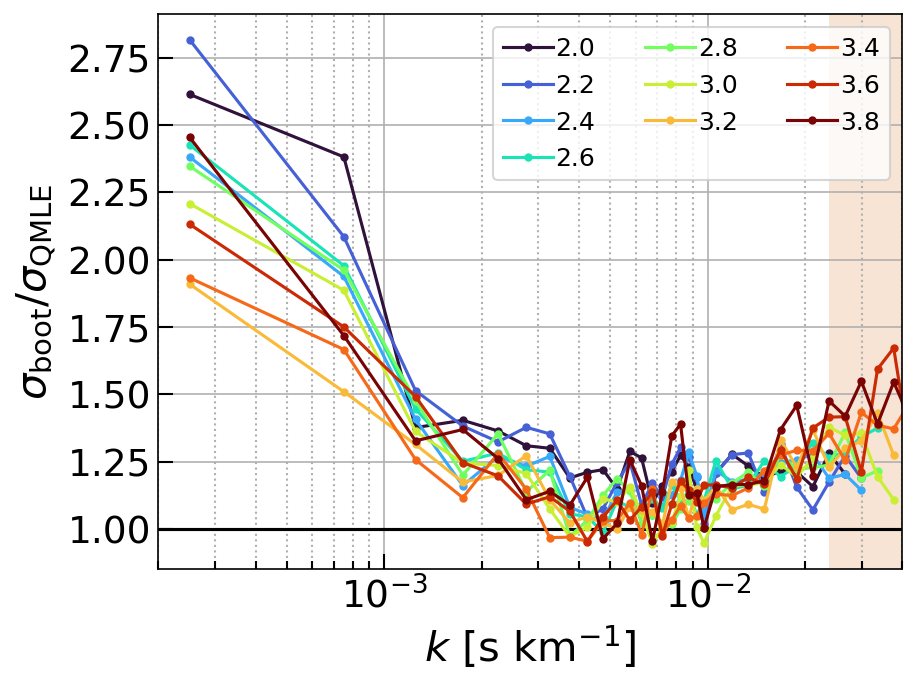}
    \caption{Ratio of the regularized bootstrap error estimates to Gaussian (QMLE) errors for \lya.
    We estimate the bootstrapped covariance matrix from $200\,000$ realizations over 256 subsamples.
    This shows that QMLE underestimates the errors on most scales and redshifts.}
    \label{fig:bootstrap_qmle_error_ratio}
\end{figure}

Figure~\ref{fig:bootstrap_qmle_error_ratio} shows that the bootstrap error estimates are mostly larger than the Gaussian estimates except between $0.003$\skm$\lesssim k \lesssim 0.1$\skm for $z\gtrsim 3.0$. QMLE consequently somewhat underestimates the errors on most scales for most redshifts.

\subsection{Side bands}
\label{subsec:sidebands}
We have investigated wavelengths larger than the \lya line in the spectrum to statistically estimate metal contamination and other systematics as they are free from neutral hydrogen absorption \citep{mcdonaldLyUpalphaForest2006, palanque-delabrouilleOnedimensionalLyalphaForest2013, chabanierOnedimensionalPowerSpectrum2019}.
As mentioned previously, we use SB 1 to estimate the metal power in the \lya forest.
In this section, we provide details for the SB 1 power spectrum measurement and further make use of the SB 2 and SB 3 regions as diagnostics of the metal power and other systematics such as noise calibration.
As before, we mask BAL features on all continuum fitting reductions, then ignore these quasars in further analysis.

We first fit the continuum in the SB 1 and SB 2 regions while fixing $\eta=1$.
We find the power in SB 1 is larger than in SB 2 as expected, except at $z=2.0$, where $P_\mathrm{SB\,2}>P_\mathrm{SB\,1}$ for $k\lesssim 0.003$\skm.
This likely points to some remaining continuum errors in the side bands.
The C~\textsc{iv} doublet feature is clearly visible in our estimates.
We refer the reader to our companion paper on modelling the doublets in the side band power spectrum \citep{karacayliFrameworkMetals2023}.

\begin{figure}
    \centering
    \includegraphics[width=\columnwidth]{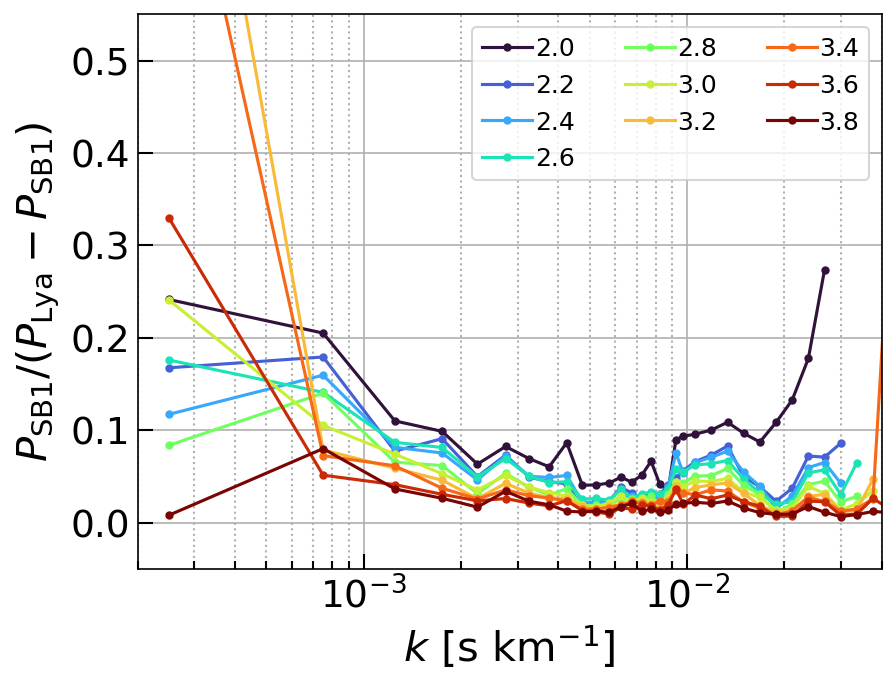}
    \caption{Ratio of metal (SB 1) power spectrum to metals-subtracted \lya power spectrum.
    The metal power is well below 20\% for $k\gtrsim 0.001$\skm, but
    it is potentially a significant source of systematic error at larger scales.
    }
    \label{fig:ratio_sb1_lyasb1subt}
\end{figure}

The accuracy of noise calibration and its dependence on SNR can also be studied using these side bands.
We divide the spectra into high ($\mathrm{SNR}>2$) and low ($\mathrm{SNR}<2$) signal-to-noise ratio samples.
This corresponds to approximately a 30/70\% split in terms of the number of quasars for both side bands.
We find that the low SNR sample sometimes has larger power, as was the case in SDSS \citep{mcdonaldLyUpalphaForest2006}, but we note that the low SNR sample yields significantly noisier \poned estimates, which hinders strong conclusions, so we further explore this dependence using variance statistics below.
Unfortunately, the power difference between the low and high SNR samples is potentially due to the SNR dependence of the pipeline noise estimates.
Therefore, we remove the metal power from the \lya forest estimates using our best estimates \emph{after} we recalibrate the pipeline noise, which is different from what was done in \citet{mcdonaldLyUpalphaForest2006}.
Finally, Figure~\ref{fig:ratio_sb1_lyasb1subt} shows the ratio of the SB 1 power spectrum to the metals-subtracted \lya power spectrum.
The metal power is potentially a large source of systematic error at $k\lesssim 0.001$\skm, but its effect is well below 20\% at higher $k$ values.

\begin{figure}
    \centering
    \includegraphics[width=\columnwidth]{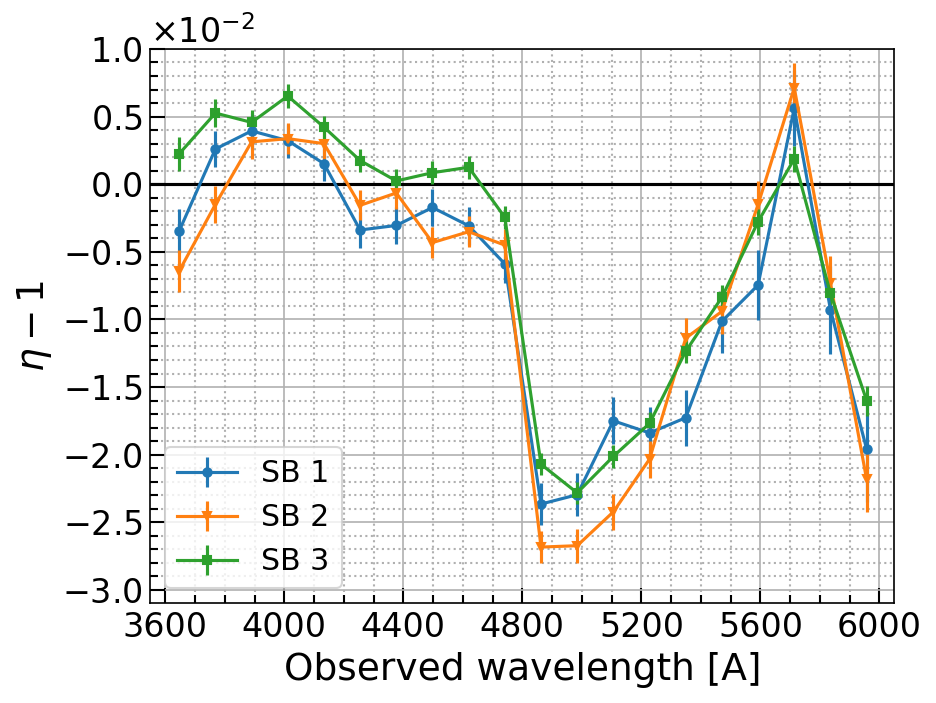} \\
    \includegraphics[width=\columnwidth]{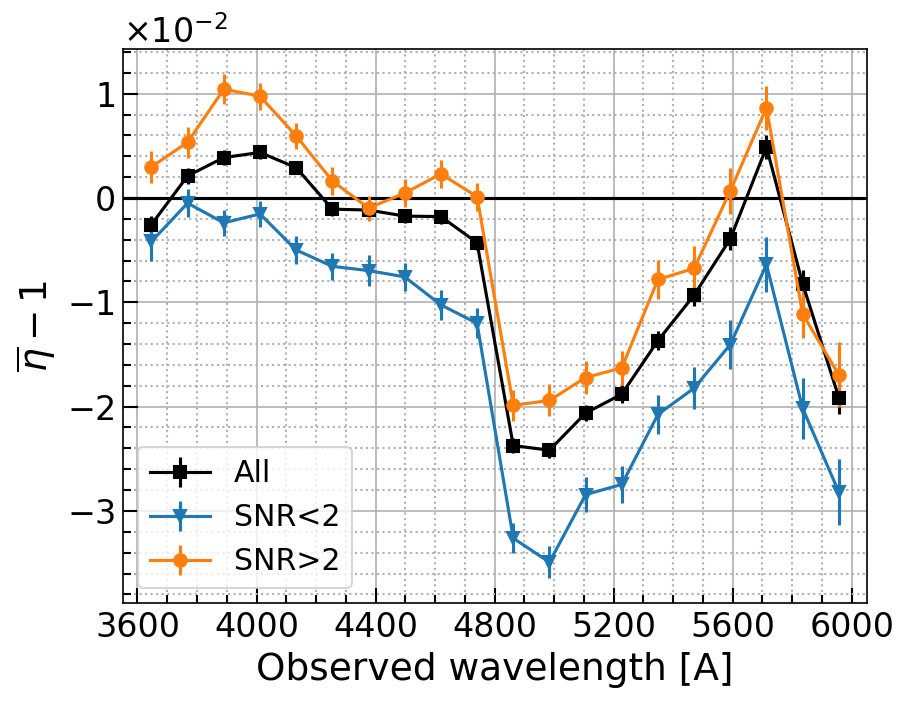}
    \caption{Pipeline noise correction term $\eta$ on all three side bands.
    These regions are relatively absorption free and can be observed in lower redshift quasars, so they provide robust statistics.
    ({\it Top}) All three side bands show the same $\eta$ trend.
    We find that the pipeline noise estimates are correct at the percent level.
    The sharp feature at 4,800~\AA\ occurs at the boundary between CCD amplifiers.
    ({\it Bottom}) Average $\eta$ over three side bands.
    The $\mathrm{SNR}>2$ sample ({\it orange circles}) has higher $\eta$ than lower $\mathrm{SNR}<2$ sample ({\it blue triangles}).
    This difference is 1.1\% on average.
    We correct the pipeline noise estimates by $\overline{\eta}$ of all spectra ({\it black squares}) and assign $1.1\%$ as our noise systematic error budget.
    }
    \label{fig:sb-three-etas}
\end{figure}

\begin{figure*}[h]
    \centering
    \includegraphics[width=0.9\linewidth]{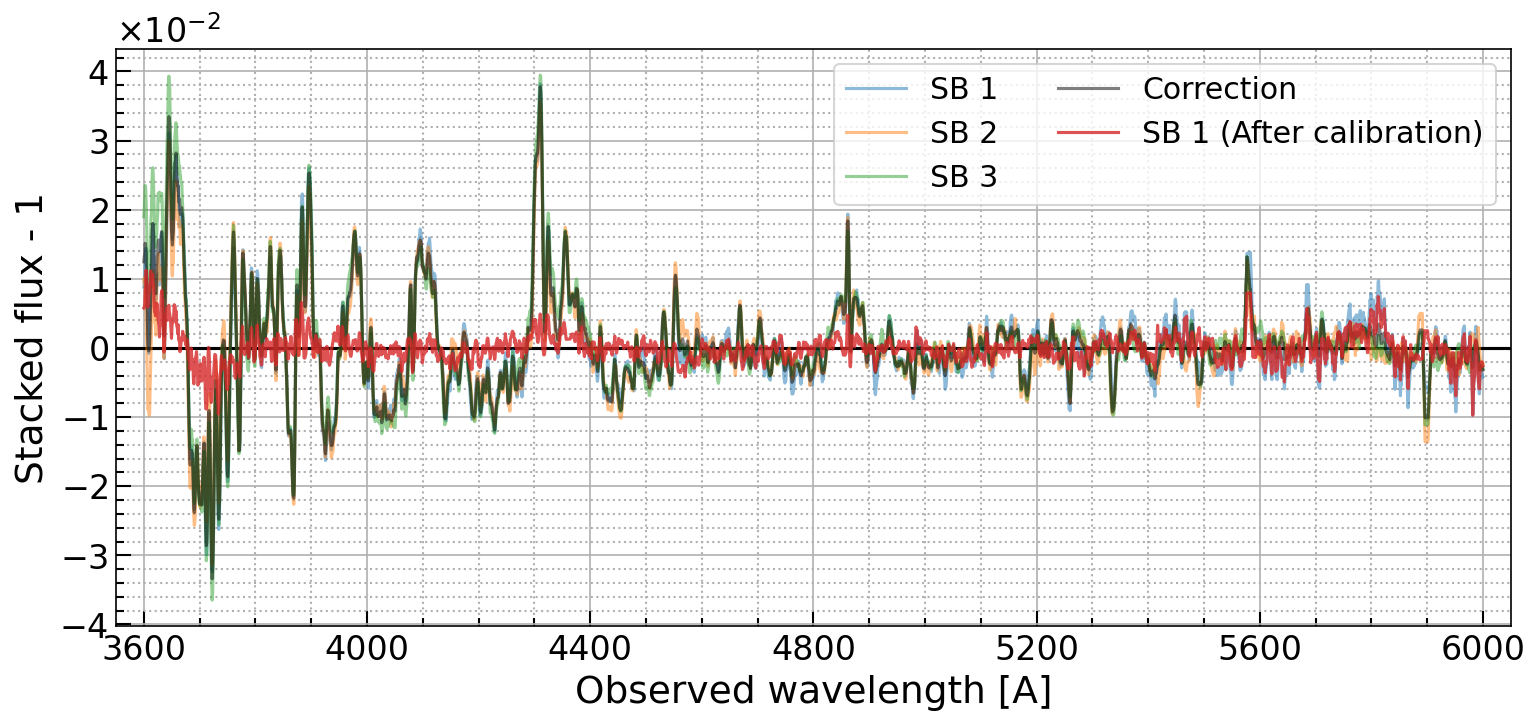}
    \caption{Stacked normalized flux from all three side bands in observed wavelength.
    We smooth the normalized flux with a 4.8\,\AA\ moving box-car average to suppress spurious fluctuations.
    Residual errors peak at most at 3\% at the Balmer and Ca~\textsc{ii} H\&K doublet lines.
    We correct the pipeline flux and noise estimates using the average of all three side bands ({\it black line}) and perform the final calibrated continuum fitting (which also includes the $\eta$ correction).
    This calibration removes significant features from SB 1 ({\it red line}), such that the remaining fluctuations are 0.2\% on average.}
    \label{fig:stacked_flux_3sbs}
\end{figure*}

\subsubsection{Noise calibration error}
As we have alluded to before, the pipeline noise estimates can suffer from miscalibrations, which then directly propagate to the final \lya \poned estimates.
Fortunately, smooth quasar continua in relatively absorption-free regions (i.e. the side bands) provide near-ideal data to investigate the estimated pipeline variance vs.\ observed variance in the data. 
These regions can furthermore be observed in numerous lower redshift quasars, so they provide robust statistical measurements.
Our continuum fitting algorithm quantifies this noise calibration error through the $\eta$ parameter, which is measured by comparing the scatter in $\delta_F$ to the reported pipeline $\sigma_\mathrm{pipe}$ values.
The pipeline noise is underestimated for $\eta>1$ and overestimated for $\eta<1$.

First, we fit the continuum on all three side bands while keeping $\eta=1$ fixed.
Then we calculate the multipoles $\langle \delta_F\rangle, \langle \delta_F^2 \rangle$ and $\langle \delta_F^4 \rangle$ in logarithmic $\sigma_\mathrm{pipe}$ bins.
We find that use of $\sqrt{(\langle \delta_F^4 \rangle-\langle \delta_F^2 \rangle^2)/N}$ as an error estimate yields biased results, so we instead estimate the error on the observed scatter $\langle \delta_F^2 \rangle$ with the delete-one Jackknife method over sub-samples.
Finally, we fit for $\eta$ and $\sigma_\mathrm{LSS}$ using equation~(\ref{eq:sigma2_cfit}), where the continuum is already taken into account in $\delta_F$.

We find that all three side bands yield similar $\eta$ values to those shown in Figure~\ref{fig:sb-three-etas}.
Our calculated $\eta$ values fluctuate by a few percent around one, which means the pipeline noise calibration is correct at the percent level.
The boundary between CCD amplifiers is responsible for the sharp feature at 4,800~\AA.
The average noise calibration correction $\overline{\eta}$ is shown in black squares in the lower panel.
We correct the pipeline noise estimates using this mean $\overline{\eta}$ as a function of wavelength and perform a final calibrated continuum fitting, which further includes the flux calibration as discussed below.
We again split the data into high ($\mathrm{SNR} > 2$) and low ($\mathrm{SNR} < 2$) signal sub-samples and measure the $\eta$ parameter in each sample.
These are shown with orange circles and blue triangles in the same lower panel of Figure~\ref{fig:sb-three-etas}.
We observe a clear SNR dependence in the noise calibration error, as has been indicated by the power spectrum estimates on these same low and high signal sub-samples.
We assign the average difference $\sigma_{\overline{\eta}} = 1.1\%$ to the systematic error budget.
\begin{equation}
    \sigma^\mathrm{syst}_N = \sigma_{\overline{\eta}} P_N.
\end{equation}

Finally, we calculate the average pipeline inverse variance as a function of flux transmission fluctuations $\delta_F$ and wavelength.
The pipeline inverse variance overweights $\delta_F=-1$ pixels at $\lambda_\mathrm{obs}=5200$\,\AA\ which we confirm does not happen in our mock analysis.
We do not observe such a peak in SB~1, so we speculate that it is a bias in noise calibration due to substantial \lya absorption.
We confirm that using $\sigma_\mathrm{LSS}$ in the continuum fitting and $P_\mathrm{fid}$ in the QMLE removes this feature from the weights.
Therefore, we conclude that it does not bias our results.

\subsubsection{Flux calibration error}
Following \citet{bourbouxCompletedSDSSIVExtended2020}, we tested possible flux calibration errors by stacking all normalized quasar spectra in the observed frame using all three side bands.
This stacked normalized flux is shown in Figure~\ref{fig:stacked_flux_3sbs}.
The residual errors are at most 3\%, and we find that the largest errors are at the locations of the Balmer series and Ca~\textsc{ii} H\&K doublet lines.

We average the stacked normalized flux in all three side bands and smooth with a 4.8\,\AA\ moving box-car average.
We then divide the flux and noise estimates by this correction factor.
This calibration removes significant features from SB 1 (red line), such that the remaining fluctuations are 0.2\% on average (smoothed on the same 4.8\,\AA\ scale).

\subsection{Systematic error budget}
\label{subsec:systematics}
We identified four sources of systematic errors.
First, given the large contribution of the noise power, our power spectrum estimates are susceptible to pipeline noise misestimates. Second, the DLA finder is not perfect and can miss some DLAs or introduce false positives, which would add power to large scales.
Third, the spectrograph resolution limits the smallest scale we can measure, and its uncertainties must be propagated to the remaining scales.
Finally, our continuum marginalization does not remove all modes of error, so we estimate the systematic error due to the remaining fluctuations.

Our systematic error analysis relies on various scalings of the underlying signal.
Directly using the power spectrum estimates introduces spurious fluctuations, so 
we instead use a smooth power spectrum $P_{\mathrm{smooth}}(k, z)$ that we calculate as follows:
We apply \texttt{SmoothBivariateSpline} to the logarithms of $1+z, k$ and \poned with corresponding statistical weights $\sigma_\mathrm{stat}/$\poned and a smoothing factor that is five times the number of data points.
This function in \textsc{scipy} is based on the algorithm presented in \citet{DierckxAlgorithmforSurface1981} and \citet{ DierckxCurveandSurface1993}.

\begin{itemize}
    \item \textbf{Noise:} As we discussed above, we use the average difference of $\eta$ values in the side bands between high and low signal sub-samples to quantify the systematic error budget shown in Figure~\ref{fig:sb-three-etas}.
        \begin{equation}
            \sigma^\mathrm{syst}_N = \sigma_{\overline{\eta}} P_N,
        \end{equation}
    where $\sigma_{\overline{\eta}} = 1.1\%$ as discussed previously. In this case, we do not smooth the noise power spectrum $P_N$ since it does not suffer from random fluctuations. Possible origins of this error include correlated CCD readout noise and errors in sky estimates per fiber. We revisit this point at the end of Section~\ref{sec:discussion}.
        
    \item \textbf{Resolution:} Based on our analysis in Section~\ref{subsec:ccd-sim}, we assign $1\%$ precision to the pipeline resolution estimates that is approximately consistent with the redshift average values in Figure~\ref{fig:ccd-image-reso}.
        \begin{equation}
            \sigma^\mathrm{syst}_{\mathrm{res}} = 1\% \times 2k^2R_z^2 \times P_{\mathrm{smooth}}(k, z),
        \end{equation}
    where $R_z\equiv c\Delta\lambda_\mathrm{DESI}/(1+z)\lambda_\mathrm{Ly\alpha}$ as defined in Section~\ref{subsec:quickquasars}. We do not directly use the values in Figure~\ref{fig:ccd-image-reso} as it could introduce double counting of noise calibration errors as spectro-perfectionism couples noise and resolution. 
        
    \item \textbf{Incomplete DLA removal:} \citet{wangDeepLearningDESIDLA2022} reports over 90\% efficiency and purity for the DLA finder for a range of SNR and column density ranges.
    To be conservative, we pick the worst performance number at SNR$=1-2$ and assign an average 15\% inefficiency ratio and multiply the smooth power estimate with the redshift average of the absolute ratio of $P_{\mathrm{wDLA}}/P_{\mathrm{true}}-1$ in Figure~\ref{fig:mock-nodla-wrt-true-ratio}.
        \begin{equation}
            \sigma^\mathrm{syst}_{\mathrm{DLA}} = 15\% \times \left\langle|P_{\mathrm{wDLA}}/P_{\mathrm{true}}-1| \right\rangle_z \times P_{\mathrm{smooth}}(k, z)
        \end{equation}
    
    \item \textbf{Continuum errors:} We use the absolute ratio of $P_{\mathrm{cont}}/P_{\mathrm{true}}$ in Figure~\ref{fig:mock-continuum-err-power}, which is on the order of $10^{-5}$.
    This error also scales with the smooth power.
        \begin{equation}
            \sigma^\mathrm{syst}_{\mathrm{cont}} = |P_{\mathrm{cont}}/P_{\mathrm{true}}|(k, z) \times P_{\mathrm{smooth}}(k, z)
        \end{equation}
    Note that continuum errors themselves are larger than $10^{-5}$ as shown in Figure~\ref{fig:meancont_varlss_mock}.
    However, our marginalization prevents the bulk of these errors from contaminating the \poned measurements
    and the remaining continuum errors are fortunately extremely small.
    Furthermore, the modes that are most affected are not in our conservative range.
\end{itemize}

\begin{figure}
    \centering
    \includegraphics[width=\columnwidth]{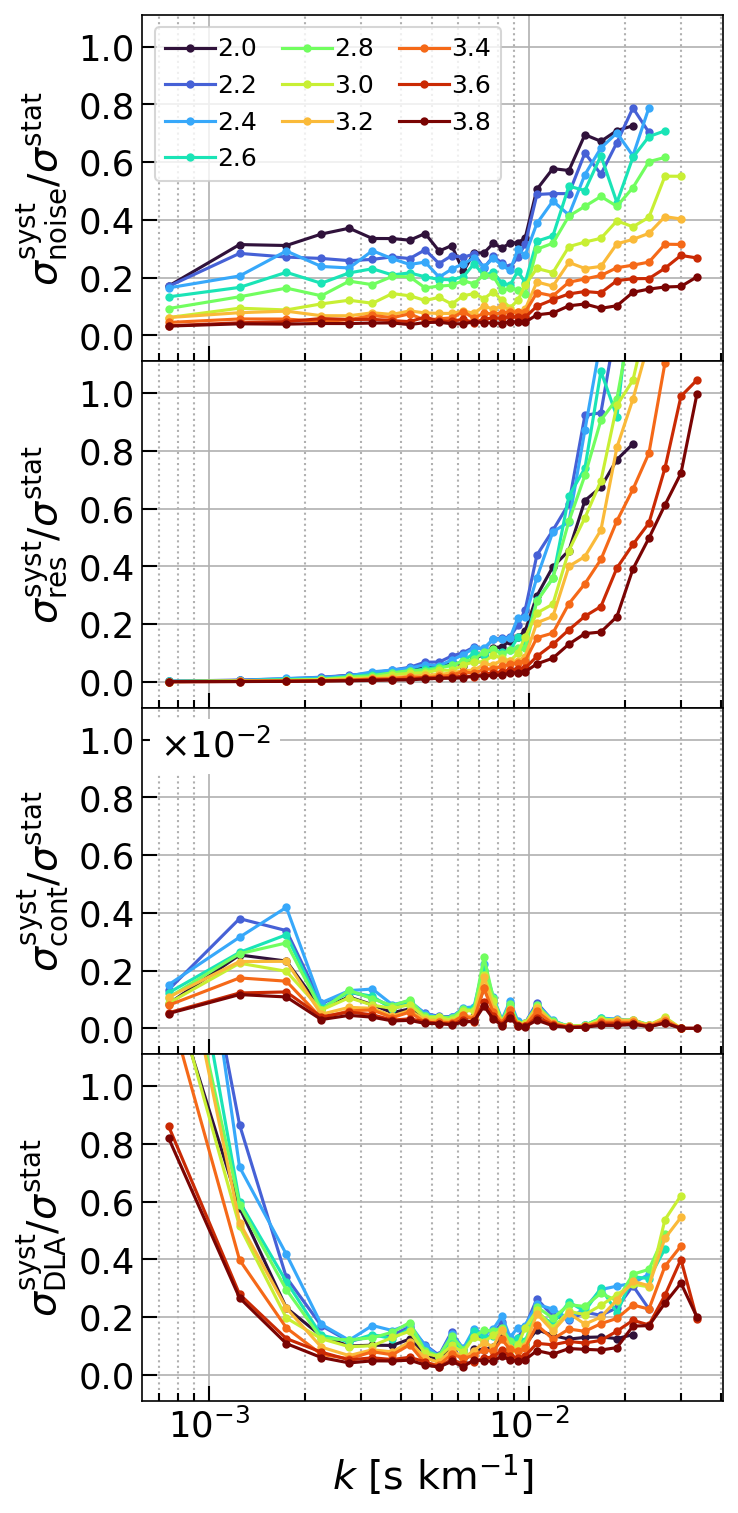}
    \caption{Systematic error budget divided by the statistical errors estimated by bootstrap analysis. Systematics due to the noise calibration is persistent on all scales and prevalent over the statistical error at lower redshifts. Resolution systematics predominantly influence small scales (high $k$).}
    \label{fig:systematic-to-stat-ratio}
\end{figure}

Figure~\ref{fig:systematic-to-stat-ratio} shows the systematic errors relative to the statistical errors. DLA systematics heavily affect the large scales, whereas the resolution systematics become relevant at $k \gtrsim 0.01$\skm (small scales). The strength of noise systematics decreases with redshift as a consequence of decreasing number of quasars. Change from linear to log-linear binning at $k=0.01$\skm increases the bin size and causes the jump at this wavenumber for the noise systematics ratio.

\section{Discussion}
\label{sec:discussion}
In this section, we first provide a qualitative comparison of our results to FFT estimates on the same DESI EDR+ data, and compare both to eBOSS measurements. We then discuss the effects of each systematics and their contribution to the covariance matrix. A full cosmological comparison with respect to physical parameters will be the subject of future analysis with DESI Year 1 data, and will require significant development to generate hydrodynamic simulations and train emulators \citep{pedersenEmulator2021, cabayolNeuralNetworkLyaEmu2023, chabanierLyaEulerianSPHsims2023}. We instead provide a comparison using a simplified version of equation~(\ref{eq:pd13_fitting_fn}).

\subsection{Consistency between methods and literature}
\begin{figure*}
    \centering
    \includegraphics[width=\linewidth]{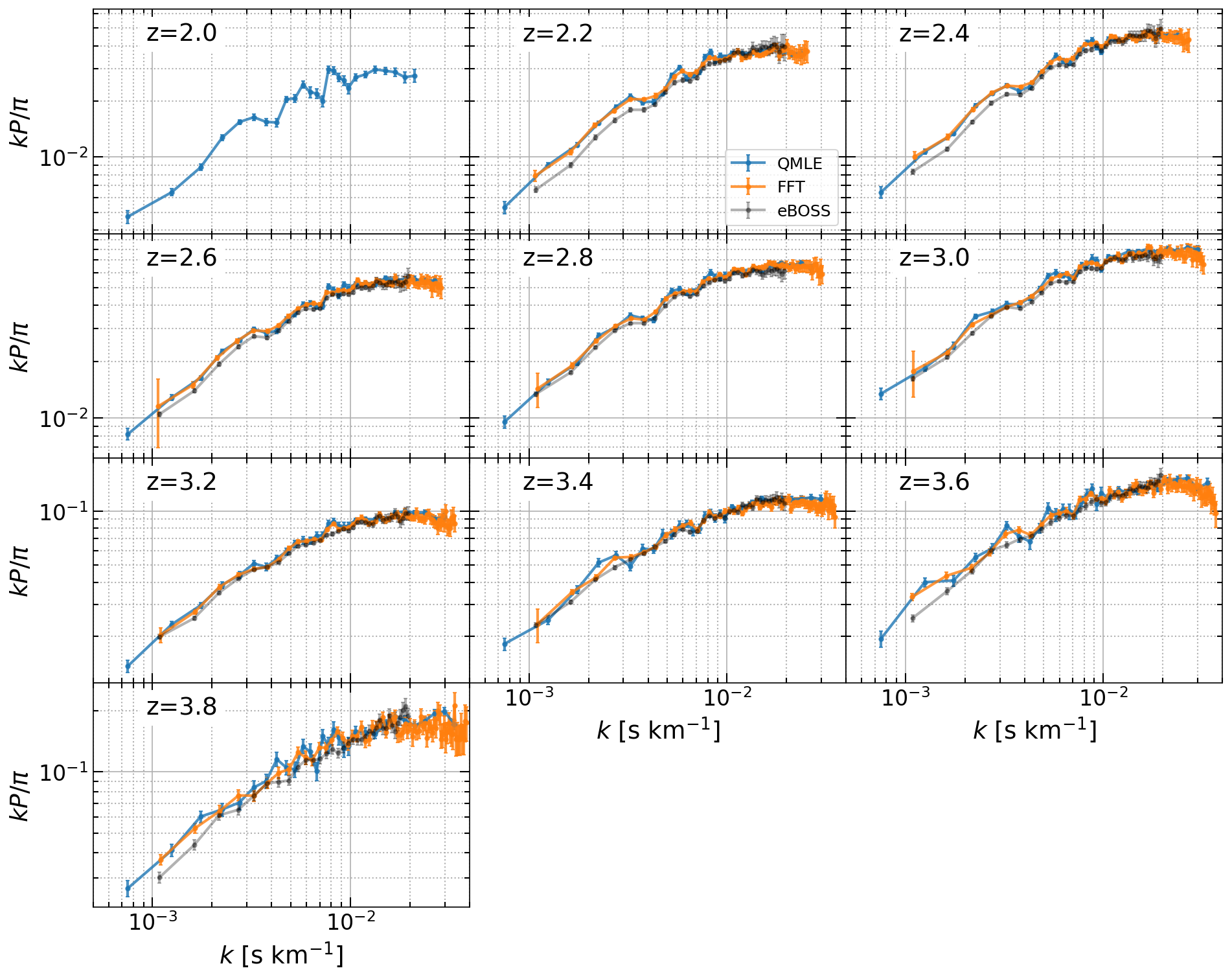}
    \caption{Comparison of \poned between FFT and QMLE methods on DESI EDR+ and to eBOSS. Both methods agree with each other overall, but disagree with eBOSS mostly at large scales at $z\lesssim 2.6$. We suspect the efficiency of the DLA finder and the noise calibration could cause this discrepancy, and we will explore this further in future work.}
    \label{fig:desi-fft-qmle-eboss-comparison}
\end{figure*}

Figure~\ref{fig:desi-fft-qmle-eboss-comparison} compares \poned from DESI EDR+ using FFT and QMLE methods.
Both methods agree with each other within error bars until half the Nyquist frequency. We note that the FFT sample is limited to SNR>1, whereas the QMLE sample is extended to SNR>0.25. Since QMLE applies a weighted average based on SNR, this choice has minimal effect, which we confirm by comparing the QMLE estimates on the SNR>1 sample. The agreement between the two estimators is noteworthy given the built-in different methods of handling noise, metals, masking, etc.
Since we find $z=2$ bin is highly contaminated by various systematics, we recommend not using our QMLE results at $z=2$ for any cosmological inference.
Specifically, we find larger power than eBOSS on large scales, and we put substantial effort into identifying the root cause. The continuum fitting method slightly differs between eBOSS and DESI analyses, which could affect these large scales.  However, our tests on mocks indicate that the effect of continuum errors on the power spectrum is small for DESI. The prime candidate then is systematics related to DLAs. We find that eliminating sub-DLAs ($\log N \lesssim 20.3$) removes power from large scales, but improvement in lower redshift bins comes with deterioration at higher redshifts. Furthermore, the DLA finder yields many false-positive, low-column density systems and is not reliable to identify such systems. The second candidate is a possible error in the noise power subtraction. However, our noise calibration correction does not fix this issue. We will study this disagreement further after additional development and testing of the DLA finder and how DLAs are included in our mock datasets, as well as with additional tests on the continuum fitting method.

\subsection{Off-diagonal systematics in the covariance matrix}
Proper use of our systematics budget requires special guidance. \citet{chabanierOnedimensionalPowerSpectrum2019} and \citet{karacayliOptimal1DLy2022} added the systematics errors in quadrature to the diagonal of the covariance matrix. This may be true if systematic uncertainties are uncorrelated between bins, but is inconsistent with our characterisation. Simply put, we characterise each systematic error by an unknown scalar $\epsilon$ multiplied with some $P(k)$ such that the uncertainty is in $\epsilon$, not in $P(k)$. For a given true signal $s(k)$, the observed data is given by $d(k) = s(k) + \epsilon P(k)$, and the covariance matrix becomes $C \equiv \langle d(k) d(k') \rangle = \langle s(k) s(k') \rangle + \langle \epsilon^2 \rangle P(k) P(k')$. In terms of the systematic error budget we presented in Section~\ref{subsec:systematics}, the covariance matrix should be modified as $\mathbf{C}_\mathrm{total} = \mathbf{C}_\mathrm{stat} + \sum_{i \in \{\mathrm{syst}\}} \bm{v}_i \bm{v}_i^T$ for each systematic error mode $v_i$. This notation also highlights the mathematical relation with the continuum marginalisation procedure. Unlike the continuum marginalization procedure which adds infinitely large error modes $v_i$ to fully remove contamination, systematic error modes are down-weighted by finite numbers.

Let us now discuss how these off-diagonal systematic error terms affect the parameters of interest. As we have noted, a full cosmological comparison is out of the scope of this work, and we instead use a fitting function based on equation~(\ref{eq:pd13_fitting_fn}). In our preliminary analysis, we found that not all parameters in equation~(\ref{eq:pd13_fitting_fn}) can be fitted properly and securely, so we simplify it by first ignoring the redshift evolution parameters $B$ and $\beta$. We instead fit each redshift bin separately and deduce the redshift evolution of each fitting parameter from these independent fits, though this redshift evolution itself is not pertinent to our discussion. We also found that the data is not sensitive enough to constrain the $k_1$ parameter and it is highly degenerate with others. These two aspects of the $k_1$ parameter destabilise the fitting, so we also eliminate it from our fitting function. We therefore instead use the following simplified fitting function:
\begin{equation}
    \label{eq:discuss} P_{\mathrm{base}}(k) = \frac{A \pi}{k_0} \left( \frac{k}{k_0} \right)^{2 +n + \alpha\ln k/k_0},
\end{equation}
where $k_{0} = 0.009$\skm as before. This formulation has simplistic yet useful characterisations such as parameter $A$ as the amplitude and $n$ as the slope of the power spectrum. Even though this procedure does not completely describe \poned, it is adequate to compare different measurements. 

Another consideration in the fitting is that the \ion{Si}{II} and \ion{Si}{III} ions absorb at wavelengths that are close to the \lya transition line \citep{mcdonaldLyUpalphaForest2006}. The more visible oscillations in the power spectrum estimates are due to the \ion{Si}{III} line at $\lambda_\mathrm{RF} = 1206.5$~\AA, which corresponds to a velocity spacing of $\Delta v = 2270$\kms. Rather than complicating our fitting function with numerous \ion{Si}{II} lines (at 1190~\AA, 1193~\AA\ and 1194.5~\AA) \citep{NIST_ASD}, we take a single, average velocity spacing of $\Delta v = 5770$\kms to model its pattern in the power spectrum. We multiply the base fitting function in equation~(\ref{eq:discuss}) with $1+a_s^2 + 2a_s\cos(k \Delta v)$ for each Si ion $s$, where $a_s$ is the relative bias with respect to neutral hydrogen \citep{mcdonaldLyUpalphaForest2006, palanque-delabrouilleOnedimensionalLyalphaForest2013}. We find the least $\chi^2$ solution using \textsc{iminuit}\footnote{\url{https://iminuit.readthedocs.io}} \citep{iminuit}, and then sample around the minimum using the \textsc{emcee}\footnote{\url{https://emcee.readthedocs.io}} sampler \citep{emcee}.

\begin{figure}
    \centering
    \includegraphics[width=\columnwidth]{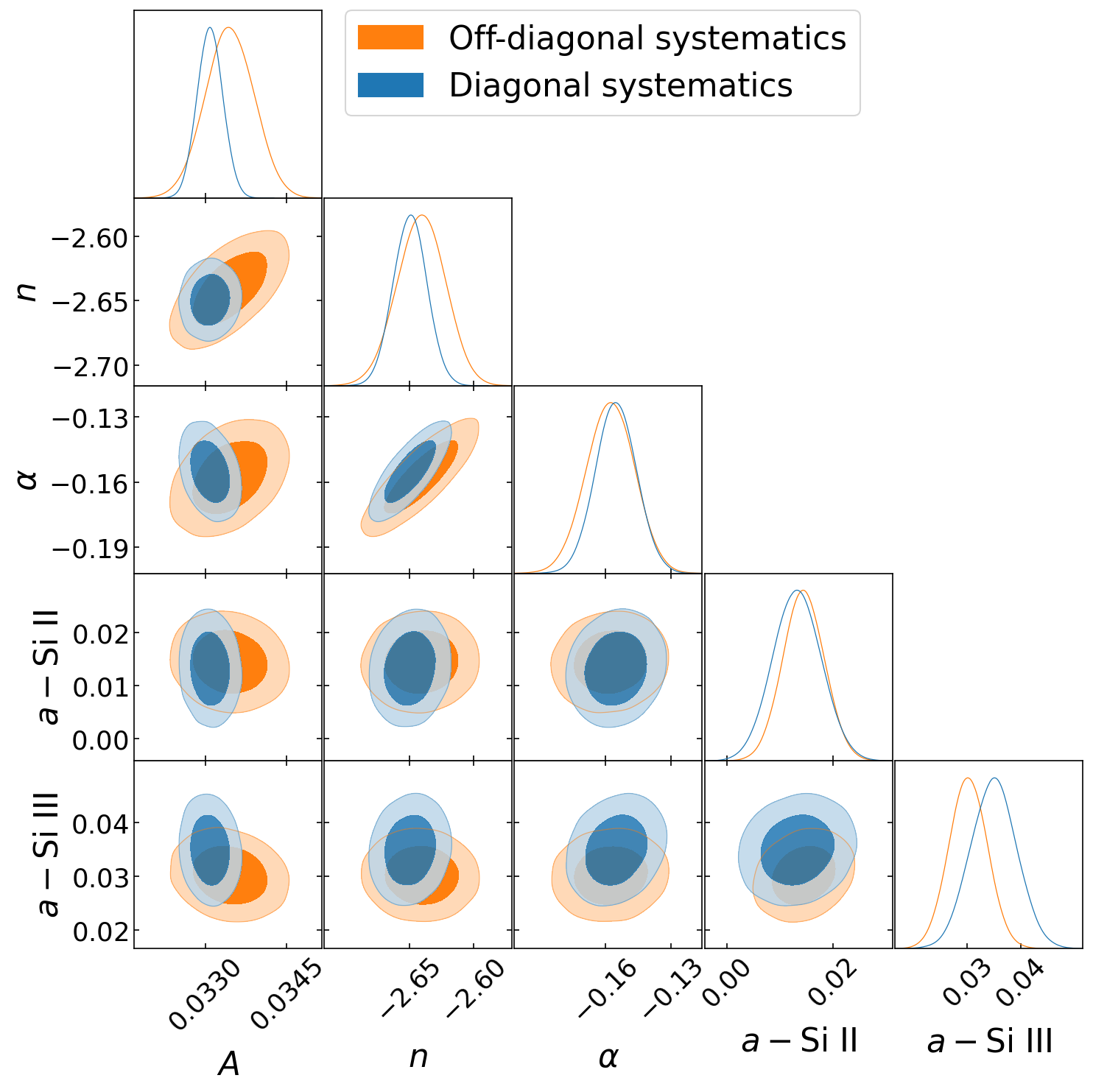}
    \caption{Best-fitting parameters and 60\% and 95\% confidence contours when the covariance matrix has systematic error contributions in its off-diagonal elements ({\it orange}) and in its diagonal only ({\it blue}) at $z=2.2$. Off-diagonal terms shift the best-fitting values and increase the error bars, which has potentially important implications for cosmological inference.}
    \label{fig:offdia-syst-corner}
\end{figure}

We apply this procedure to our metal-power subtracted power spectrum results. The covariance matrices are obtained with the bootstrap method. Figure~\ref{fig:offdia-syst-corner} shows a corner plot at redshift $z=2.2$ at 68\% and 95\% confidence levels generated using the \textsc{getdist} package \citep{getdist}. Results from adding the systematic error budget to the diagonals of the covariance matrix are shown in blue contours. Our recipe for the off-diagonal contribution (shown in orange) not only enlarges the confidence area (i.e. increases the error bars), but also shifts the best-fitting parameter value, which is particularly pronounced in the amplitude parameter $A$. This has potentially important implications for cosmological inference. These off-diagonal covariance matrix contributions will be important to incorporate into future analyses.

\begin{table}
    \centering
    \begin{tabular}{l|r|r|r}
        & \multicolumn{3}{c}{Increase in error} \\
        Systematics & $A$ & $n$ & $\alpha$ \\
        \hline
        Noise & 49.8\% & 6.9\% & 0.7\% \\
        Resolution & 34.0\% & 89.1\% & 26.5\% \\
        DLA & 0.9\% & 2.8\% & 30.5\% \\
        \hline
        All & 74.3\% & 99.9\% & 62.5\%
    \end{tabular}
    \caption{Percentage increase in error given by the minimizer for each systematics at $z=2.8$. The precision of the amplitude $A$ is nearly equally affected by noise and resolution systematics, whereas for $n$, it is thoroughly affected by resolution systematics.}
    \label{tab:increase_in_error_syst}
\end{table}

\subsection{Effects on parameters of each systematic} 
\begin{figure}
    \centering
    \includegraphics[width=\columnwidth]{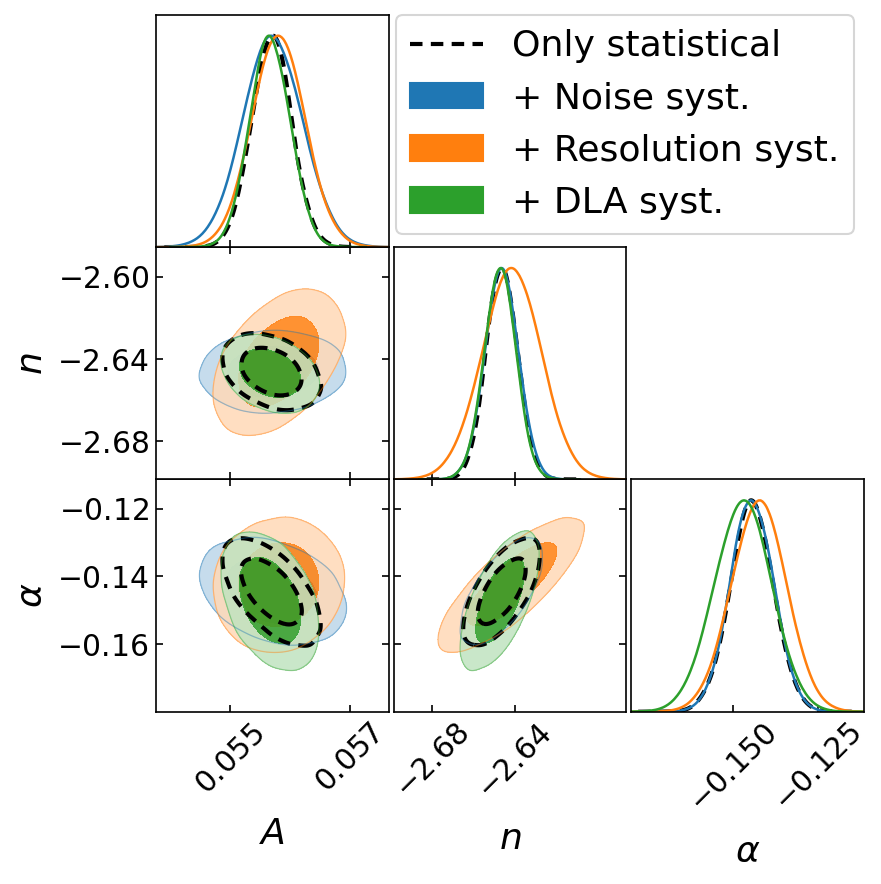}
    \caption{The 60\% and 95\% confidence regions using only the statistical covariance ({\it black dashed lines}), and individually adding the noise systematics ({\it blue}), resolution systematics ({\it orange}) and DLA systematics ({\it green}) at $z=2.8$. DLA systematics do not seem to increase the error bars as much as the other sources of systematic error. The noise and resolution systematics visibly enlarge the contours, and the resolution systematics further shifts the best-fitting values.}
    \label{fig:each-syst-corner}
\end{figure}

The systematic error budget diminishes the benefit of having a large and statistically powerful data set. As the data size increases, the statistical error bars decrease, but the systematic error budget stays the same and eventually saturates the constraining power of any analysis. Using the same off-diagonal recipe for the covariance matrix, we break down how each systematic influences the error bar of inferred parameters. First starting with only the statistical covariance matrix from bootstraps, we add each systematic individually to the covariance matrix, find the least $\chi^2$ solution, and sample around this minimum. Figure~\ref{fig:each-syst-corner} shows the confidence regions using the statistical covariance in black dashed lines, where the noise systematics are in blue, the resolution systematics are in orange, and the DLA systematics are in green at $z=2.8$. The noise and resolution systematics visibly enlarge the contours, and resolution systematics further shift the best-fitting values. The DLA systematics do not seem to increase the error bars significantly compared to the other two. Based on our analysis with the minimizer, we find that the error in the $A$ parameter increases by 50\% due to noise systematics, 34\% due to resolution systematics, and 74\% when all systematics are included. However, the slope parameter $n$ is affected more unevenly: its error increases only 7\% due to noise systematics, but 89\% due to resolution systematics. Table~\ref{tab:increase_in_error_syst} and Table~\ref{tab:best-syst-decomp-numbers} detail these numbers.

\begin{table}
    \centering
    \begin{tabular}{*{6}{l}}
         & Best & $\pm$ (stat.) & $\pm$ (noise) & $\pm$ (reso.) & $\pm$ (DLA) \\
         \hline
        $A$ & 0.05578 & 0.00033 & 0.00037 & 0.00030 & 0.00005 \\
        $n$ & -2.643 & 0.008 & 0.003 & 0.012 & 0.002 \\
        $\alpha$ & -0.1451 & 0.0064 & 0.0008 & 0.0050 & 0.0054
    \end{tabular}
    \caption{Best-fitting values and error estimates given by the minimizer for each systematics at $z=2.8$.}
    \label{tab:best-syst-decomp-numbers}
\end{table}

Year 1 DESI spectra will provide even more statistical power. To fully exploit its statistical power, we will have to prioritise mitigation of the noise calibration and spectrograph resolution systematics. 
As mentioned in the previous section, correlated CCD readout noise and sky subtraction errors can source noise systematics. The small number of quasars in the EDR+ sample limited our ability to perform ambitious data splits. The first-year data will have about a million quasars, which will enable a granular investigation. To remedy the noise systematics, we will first split the data into ten spectrograph subsamples, and then further subdivide the data within each spectrograph into two to isolate different CCD amplifiers. This division results in about 50,000 quasars in each region, approximating the sample size in this analysis, and so should provide enough statistics to study \poned\ in each subsample precisely. CCD image simulations could also be studied for noise recalibration parameters as well as refining the pipeline resolution estimates.

\section{Summary}
\label{sec:summary}
The 1D \lya forest power spectrum \poned quantifies the clustering of diffuse neutral hydrogen gas in the intergalactic medium.
\poned has been measured from various data sets, and has been used to constrain the thermal state of the IGM, the sum of neutrino masses, and various dark matter models.
DESI will collect over 700~000 $z>2$ quasars during the five-year survey and will provide enormous statistical power for future \poned cosmology. It will be able to measure \poned from $z=2$ to $z=5$ and from approximate scales of 60~Mpc down to 1~Mpc.

This power must be matched with rigorous studies of data and estimation methods.
In this work, we employed the quadratic maximum likelihood estimator (QMLE) to measure \poned.
QMLE is built to be statistically optimal and robust against \lya forest-specific challenges such as gaps in spectra and errors due to continuum fitting.
Additionally, QMLE benefits from the resolution matrix output of spectro-perfectionism that preserves the full 2D resolution of the spectrograph.

In order to test the pipeline at various stages, DESI collected thousands of spectra during its Survey Validation phase.
We used these early spectra to determine if the DESI \poned pipeline (such as noise calibration, resolution matrix, continuum fitting and \poned estimator itself) is accurate at the desired level and measure the initial \poned from DESI.

The quasar continuum estimation is a potential source of large-scale uncertainty and bias.
We described each quasar with two free fitting parameters (amplitude and slope) that multiply a mean continuum, and fit each continuum in the forest region.
The two-parameter description is simplistic in terms of quasar continuum diversity; and furthermore fitting the continuum in the forest region removes the large-scale density information.
Using synthetic spectra, we found that the estimated large-scale variance matches the input, but the estimated mean continuum deviated from the truth.
Even though the estimated mean continuum was different, we showed that QMLE successfully marginalized out large-scale biases, and the residual power due to unmarginalized continuum error modes was significantly smaller than the signal.


The small-scale structure in the \lya forest is extremely important for the many science applications of \poned, and accurate knowledge of the spectrograph resolution is important to make the best use of the smallest scales. 
We created CCD image simulations with approximately 45~000 $z>2.1$ quasars (ten simulated observations of just quasars) and extracted the 1D spectra of these quasars using the DESI pipeline.
We used these simulations to demonstrate that the resolution matrix produced by the pipeline was valid and did not require any modifications.
We also showed that the PSF is well approximated by the pipeline's PSF model and that systematic errors due to PSF mismatch between the true PSF and the PSF fit by the pipeline were consistent with 1\% precision on the resolution.

The noise power subtraction is not an insignificant part of the \poned estimation for DESI's medium-SNR spectra.
Hence, any miscalibration of the pipeline noise is directly transmitted to the final results.
We showed that the side band power subtraction that removes the metal power also eliminates some of this miscalibration, although not perfectly.
Therefore, we examined the variance statistics of the transmission fluctuations in the relatively absorption-free side band regions to quantify the noise and flux miscalibration of the pipeline.
We showed that these calibration errors are small (around a few percent) and corrected them in our analysis.
We also found that the pipeline noise estimates are still coupled to the signal, and the noise calibration error depends on SNR.
We do not try to correct for this SNR dependence and instead include this effect in the systematic error budget.

In order to have accurate error estimates, we relied on bootstrap realizations, which yielded larger statistical errors than their Gaussian counterparts from QMLE on almost all scales and redshifts.
We further identified and quantified four major systematic error sources: noise, resolution, incomplete DLA removal, and continuum errors.
Off-diagonal terms in the covariance matrix due to these systematic errors enlarge the error bars of inferred parameters and can also shift the best-fitting values, hence biasing cosmological interpretations. We showed that the noise calibration and resolution systematics weaken the statistical power of the current DESI early data, and will be priorities for additional study and mitigation for the Year 1 analysis. For the latter, we plan to conduct data split studies to quantify and mitigate noise systematics, create more extensive image simulations to quantify resolution systematics, and produce more synthetic datasets to better quantify the performance of the DLA finder and its impact on our results.  

We found that the resulting \poned measurement from DESI early data and two months of main survey was in agreement between QMLE and FFT results, which is remarkable since the two estimators have different approaches for systematics. Furthermore, the two methods apply different SNR thresholds to the DESI sample. We confirmed that QMLE results are not affected by this threshold, which is expected since QMLE applies a weighted average based on SNR. 
These DESI QMLE and FFT \poned\ results are $5-15\%$ larger than previous measurements from eBOSS and in $1.5-3\sigma$ tension with those results.
We investigated: 1) DLA finder efficiency and column density cuts; 2) noise calibration errors to explain this disagreement.
However, none of these investigations offered a satisfactory explanation for the disagreement with eBOSS.
This feature is worth further attention in future work.

To conclude, the DESI spectral pipeline works exceptionally well.  The major analysis pipeline errors are no larger than few percent, and QMLE is well suited for DESI \poned measurements.
The next five years will bring incredible accuracy, precision, and power to \lya forest cosmology.

\section*{Acknowledgments}
\label{sec:acknow}
This material is based upon work supported by the U.S. Department of Energy (DOE), Office of Science, Office of High-Energy Physics, under Contract No. DE–AC02–05CH11231, and by the National Energy Research Scientific Computing Center, a DOE Office of Science User Facility under the same contract. Additional support for DESI was provided by the U.S. National Science Foundation (NSF), Division of Astronomical Sciences under Contract No. AST-0950945 to the NSF’s National Optical-Infrared Astronomy Research Laboratory; the Science and Technology Facilities Council of the United Kingdom; the Gordon and Betty Moore Foundation; the Heising-Simons Foundation; the French Alternative Energies and Atomic Energy Commission (CEA); the National Council of Science and Technology of Mexico (CONACYT); the Ministry of Science and Innovation of Spain (MICINN), and by the DESI Member Institutions: \url{https://www.desi.lbl.gov/collaborating-institutions}. Any opinions, findings, and conclusions or recommendations expressed in this material are those of the author(s) and do not necessarily reflect the views of the U. S. National Science Foundation, the U. S. Department of Energy, or any of the listed funding agencies.

The authors are honoured to be permitted to conduct scientific research on Iolkam Du’ag (Kitt Peak), a mountain with particular significance to the Tohono O’odham Nation.

\section*{Data availability}
All data points shown in the figures are available in simple text files on the following website: \url{https://doi.org/10.5281/zenodo.8007370}.
Some underlying DESI spectra will be publicly available as early data release (EDR) in 2023.
We added two months of main surveys to improve our statistics.
The main survey spectra will be made publicly available as part of Year 1 DR in the future.

\bibliographystyle{mnras}
\bibliography{p1d-refs.bib}

\appendix
\section{Continuum error marginalization}
\label{app:continuum_marg}
We fit the quasar continuum in the forest region as discussed previously.
This method not only fits for the quasar continuum (and the mean transmission of the IGM), but also fits the large-scale density modes in the forest.
This removes the large-scale information from the estimated flux transmission fluctuations and further cause distortions in the 3D correlation function \citep{slosarLya3DBOSS2011, bautistajuliane.MeasurementBaryonAcoustic2017, bourbouxCompletedSDSSIVExtended2020}.
Fortunately, the modes of error are known; and the QMLE is capable of marginalizing out these errors \citep[Appendix B]{slosarMeasurementBaryonAcoustic2013}. 

The continuum marginalization in \citet{karacayliOptimal1DLy2020} is implemented by modifying the covariance matrix to $\mathbf{C}' = \mathbf{C} + N \bm v \bm v^{\mathrm{T}}$, where $N$ is large and $\bm v$ is the mode to be marginalized out. Then, one can show that $\mathbf{C}'^{-1}(\bm{\delta}_F' + \alpha \bm v) \approx \mathbf{C}^{-1} \bm{\delta}_F'$, where the new data vector $\bm{\delta}_F'$ is orthogonal to $\bm v$, which effectively removes any information from data that is in mode $\bm v$.

We updated our continuum marginalization technique for more numerical stability.
Instead of adding large numbers to the covariance matrix, which could destabilize the inversion, we take advantage of the Sherman-Morrison formula \citep{shermanmorrison1950}:
\begin{equation}
    (\mathbf{C}+\bm{v}\bm{v}^\mathrm{T})^{-1} = 
    \mathbf{C}^{-1} 
    - \frac{\mathbf{C}^{-1} \bm{v} \bm{v}^\mathrm{T} \mathbf{C}^{-1}}
    {1 + \bm{v}^\mathrm{T} \mathbf{C}^{-1} \bm{v}}.
\end{equation}
Since the covariance matrix $\mathbf{C}$ is symmetric, this formula can be calculated by defining an intermediate vector $\bm{y} = \mathbf{C}^{-1} \bm{v}$. 
Moreover, the marginalization vector mode $\bm{v}$ is theoretically multiplied by a large number, such that $\bm{v}^\mathrm{T} \mathbf{C}^{-1} \bm{v} \gg 1$,
and that large number cancels out by the division, hence does not need to be explicitly defined anymore. 
Putting these together also makes the workings of the marginalization clearer:
\begin{align}
    \mathbf{C'}^{-1} & = \mathbf{C}^{-1} 
    - \frac{\bm{y} \bm{y}^\mathrm{T}} {\bm{y}^\mathrm{T} \bm{v}} \\
    \mathbf{C'}^{-1} \bm{v} &= \mathbf{C}^{-1}\bm{v} - \frac{\bm{y} \bm{y}^\mathrm{T}\bm{v}} {\bm{y}^\mathrm{T} \bm{v}} = 0
\end{align}

There are two caveats to this approach. 
First, when we marginalize over multiple templates, we have to iterate this formula with the updated covariance matrix in each step.
Fortunately, this does not introduce additional memory strain in our case since we already store the inverse covariance matrix in memory, but other applications may prefer the Woodbury formula \citep{woodbury1950inverting}.
Second, trying to remove two templates that are approximately the same results in a division by near zero.
After removing the first vector $\bm{v}_1$, the second vector results in $\bm{y}_2 = \mathbf{C'}^{-1} \bm{v}_2 \approx 0$.
To prevent numerical problems, we first store all $n$ template vectors in a rectangular matrix and perform a singular value decomposition.
This yields an orthogonal basis for the templates and vectors with small singular values can be left out of the marginalization process.

\section{Signal-noise coupling and smoothing}
\label{app:signal_noise_coupling}
Another important caveat from our preliminary analysis on mocks is the well-known signal-noise coupling in the \lya forest.
The pipeline noise is coupled to the observed flux by construction through Poisson statistics, which means pixels with less flux are weighted more in the quadratic estimator formalism \citep{mcdonaldLyUpalphaForest2006}.
However, this does not mean the noise estimates are wrong, the noise is non-diagonal, or there are cross-correlations between signal and noise.
The simulations truthfully populate the mocks with noise using independent random numbers.
The core problem is that these (correct) noise amplitudes are biased weights for the inverse variance weighted averages since they depend upon the signal (see below for an analytical description of this effect).
To mitigate this problem, the DESI pipeline determines a smoothed version of the sky-subtracted spectrum of each target that is obtained with a convolution using a Gaussian kernel of $\sigma=10$\AA\; \citep{guySpectroscopicDataProcessingPipeline2022}.
However, our preliminary analysis showed this is not enough to fully uncouple the signal and noise, so we further smooth out noise estimates in only their contribution to the covariance matrix in QMLE.

First, we find the median $\mathcal{M}(x)$ and the median absolute deviation $\mathcal{D}(x) = \mathcal{M}(|x_i - \mathcal{M}(x)|)$ in the pipeline noise estimates while ignoring pixels with $\sigma_\mathrm{pipe}>1000$.
Noise outliers are then identified by
\begin{equation}
    \sigma_\mathrm{pipe} > \mathcal{M}(\sigma_\mathrm{pipe}) + 3.5\times\mathcal{D}(\sigma_\mathrm{pipe}),
\end{equation}
and replaced with the median pipeline noise value.
We pad the noise array by 25 pixels at both ends with the edge values to mitigate any boundary effects.
We convolve the resulting noise estimates by a hybrid Gaussian box-car window function, where
the smoothing kernel has a size of 51 pixels and a Gaussian sigma of 20 pixels.
After applying this hybrid boxcar-Gaussian smoothing kernel, we return outlier values to their original positions.
This smoothing broadly captures the spectrograph behaviour while still down-weighting the masked or high-variance pixels.

\textbf{Analytical expressions:} The pipeline noise estimate of a CCD pixel depends on the exposure time $t_{\mathrm{exp}}$, and has the following contributions:
\begin{itemize}
    \item Source electrons $N_\mathrm{source}=C_q(\lambda) F(\lambda) \propto t_{\mathrm{exp}}$. Note $F = \overline{F} (1+\delta)$
    \item Sky electrons $N_{\mathrm{sky}} \propto t_{\mathrm{exp}}$.
    \item Dark electrons $\propto t_{\mathrm{exp}}$. We can absorb this term into sky contribution.
    \item Read noise electrons as Gaussian noise with constant $\sigma_G^2$ independent of $t_{\mathrm{exp}}$.
\end{itemize}
Then, pipeline variance estimate on flux $f$ is $\sigma^2_{f, \mathrm{pipe}} = N_\mathrm{source} + N_\mathrm{sky} + \sigma^2_G$.
We can write the pipeline noise estimate on $F$ as follows:
\begin{equation}
    \sigma^2_{F, \mathrm{pipe}} = \frac{\sigma^2_{f, \mathrm{pipe}}}{C_q^2(\lambda)} = \frac{F}{C_q} + \frac{N_\mathrm{sky} + \sigma^2_G}{ C^2_q}.
\end{equation}

The total variance includes large-scale structure fluctuations $\sigma^2_\mathrm{LSS}$, which should be multiplied by the mean IGM flux $\overline{F}$.
\begin{equation}
    \sigma^2_{F} = \overline{F}^2\sigma^2_\mathrm{LSS} +  \frac{F}{C_q} + \frac{N_\mathrm{sky} + \sigma^2_G}{ C^2_q}.
\end{equation}
Let us write this variance on $F$ as follows:
\begin{align}
    \sigma_F^2 &= \frac{1}{C_q} \left[ F + C_q  \overline{F}^2 \sigma^2_\mathrm{LSS} + \frac{N_\mathrm{sky} + \sigma^2_G}{ C_q} \right] \\
    \sigma_F^2 &= \frac{F + \Lambda}{C_q},
\end{align}
where we defined
\begin{equation}
    \Lambda \equiv C_q  \overline{F}^2 \sigma^2_\mathrm{LSS} + \frac{N_\mathrm{sky} + \sigma^2_G}{ C_q}.
\end{equation}
Using inverse variance weights means $w_i \propto (F+\Lambda)^{-1}$. Since $0<F<1$, our weights are highly correlated with the signal if $\Lambda \lesssim 1$. Unbiased limit $\Lambda \gg 1$ is satisfied when the sky dominates the signal (noisy spectra) or the quasar is so bright that the LSS term dominates the variance.

\begin{figure}
    \centering
    \includegraphics[width=\columnwidth]{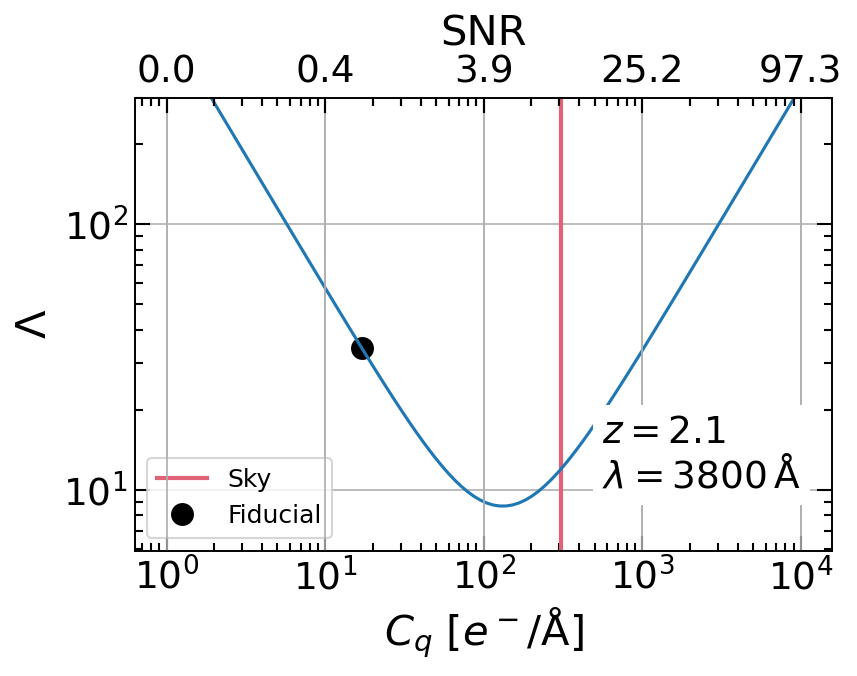}
    \caption{Coupling parameter $\Lambda$ vs quasar brightness $C_q$ as electrons per Angstrom for DESI fiducial values at 3800\AA.
    The upper axis is the human-readable SNR of the quasar.
    DESI expects mean SNR=0.45, which corresponds to $\Lambda=34$ (low coupling).}
    \label{fig:continuum-vs-Lambda}
\end{figure}

There are two important points to note. 
First, strictly speaking, pipeline noise estimates \textbf{are} correct.
The problem is not an error in the pipeline but in the nature of \lya forest.
As we mentioned in the main text, DESI decouples signal and noise by smoothing source contribution at 10\AA\, scale \citep{guySpectroscopicDataProcessingPipeline2022}.
Second, noise \textbf{is} still diagonal, and there are \textbf{no} cross correlations between signal and noise.
Noise on each pixel is another independently generated random number.
Therefore, even though its amplitude depends on the signal, this dependence does not introduce auto or cross correlations to noise.

DESI already has expected values for these quantities. 
Figure~\ref{fig:continuum-vs-Lambda} shows coupling parameter $\Lambda$ vs quasar SNR.
Low SNR quasars manifest less coupling (high $\Lambda$) since they are dominated by sky and read noise.
High SNR quasars also manifest less coupling as the LSS term dominates.
On average, DESI expects the mean quasar SNR to be 0.45, which corresponds to $\Lambda=34$.

The weighted average estimator for the mean flux is given by $\hat{\overline{F}} = \sum w_i F_i/\sum w_i$, where $w_i = (F_i+\Lambda)^{-1}$, where $F$ is signal only as before.
The expected value of this estimator can be calculated by using the probability distribution function $\mathcal{P}(F)$.
\begin{align}
    \left\langle \hat{\overline{F}} \right\rangle &= A \int_0^1 \frac{F}{F+\Lambda} \mathcal{P}(F) \ddif F \\
    A^{-1} &= \int_0^1 \frac{1}{F+\Lambda} \mathcal{P}(F) \ddif F
\end{align}
We can Taylor expand these expressions for $\Lambda\gg 1$. The normalisation is $A \approx \Lambda \left(1+\overline{F}/\Lambda\right)$.
The mean flux estimate is
\begin{align}
    \langle \hat{ \overline{F}} \rangle &= A \int_0^1 \frac{F}{F+\Lambda} \mathcal{P}(F) \ddif F \\
    &\approx \frac{A}{\Lambda} \int F \left( 1-\frac{F}{\Lambda} \right) \mathcal{P}(F) \ddif F \\
    &= \frac{A}{\Lambda} \left(\overline{F} - \frac{\overline{F^2}}{\Lambda} \right) = \left(1+\frac{\overline{F}}{\Lambda}\right) \left(\overline{F} - \frac{\overline{F^2}}{\Lambda} \right)
\end{align}
Finally, the error on the mean flux estimate is
\begin{equation}
    \Delta\overline{F} = - \frac{\overline{F}^2 \sigma^2_\mathrm{LSS}}{\Lambda}.
\end{equation}

The same calculation can be done for two-point statistics.
Gauss–Hermite quadrature is still the best way to numerically calculate the expected value.
We can either use the true mean flux or use the biased mean flux estimates from weighted averages to calculate $\delta$s.
It is worth noting that using biased mean flux does not scale but shifts estimated deltas.
\begin{equation}
    \delta_i \rightarrow \frac{\overline{F}(1+\delta_i)}{\overline{F} + \Delta\overline{F}} -1 = \delta_i - \frac{\Delta\overline{F}}{\overline{F}} 
\end{equation}
This does not make a big difference and only affects the $k=0$ mode of the power spectrum.
However, the correlation function asymptotically approaches a constant at large scales.
Assuming true deltas are used with biased weights, the error on the two-point function estimate is
\begin{equation}
    \Delta \xi_{ij} = - \frac{\overline{F}}{\Lambda} \left(\langle \delta_i^2\delta_j \rangle+ \langle\delta_i\delta_j^2\rangle\right).
\end{equation}
In other words, estimated two-point statistics are contaminated by three-point statistics and distorted at all scales through correlations between weights.

\begin{figure}
    \centering
    \includegraphics[width=\columnwidth]{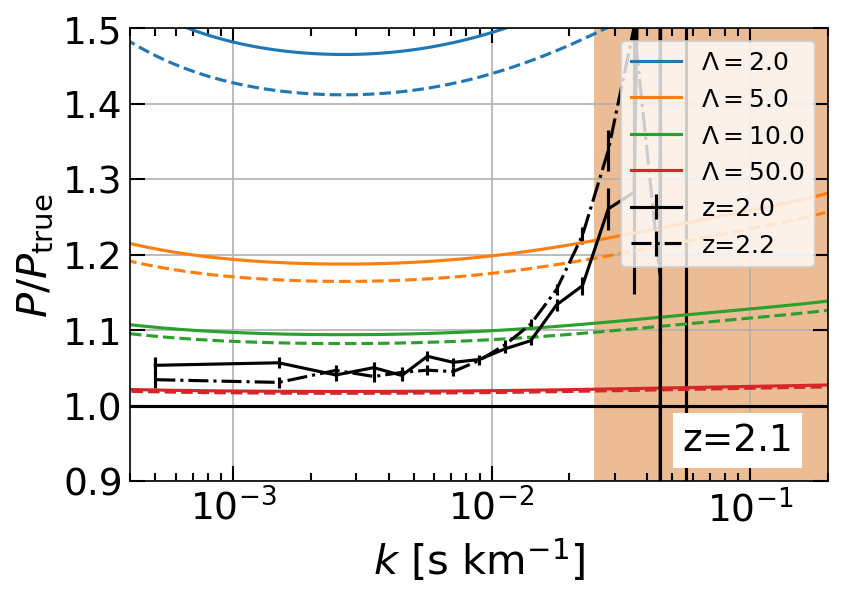}
    \caption{Biased power spectrum estimates for different $\Lambda$ values.  Using true mean flux to extract deltas ({\it dashed lines}) shows slightly less bias than using biased delta extraction ({\it solid lines}).
For comparison, our estimates from mocks without smoothing fall close to $\Lambda=50$ ({\it red line}) as expected from the fiducial value.}
    \label{fig:powerspectrum-measure-Lambda}
\end{figure}

We have numerically calculated both expressions.
For brevity, Figure~\ref{fig:powerspectrum-measure-Lambda} shows only biased \poned with respect to different $\Lambda$ values.
Using true mean flux to extract deltas (dashed lines) shows slightly less bias than using biased delta extraction (solid lines).
For comparison, our estimates from mocks without smoothing fall close to $\Lambda=50$ (red line) as expected from the fiducial value.
As we note in the main text, smoothing the noise estimates solves this problem.

\bsp	
\label{lastpage}
\end{document}